\documentclass[structabstract]{aa}
\pdfoutput=1

\usepackage[latin5]{inputenc}

\usepackage{graphicx}
\usepackage{txfonts}
\usepackage{color}
\usepackage{lscape}
\usepackage{natbib}

\newcommand{\farcss}{\mbox{\rlap{.}$''$}}

\newcommand{\kms}{km~s$^{-1}$}
\newcommand{\vlsr}{V$_\mathrm{LSR}$}
\newcommand{\vhel}{V$_\mathrm{hel}$}
\newcommand{\msun}{\mbox{M$_{\odot}$}}
\newcommand{\lsun}{\mbox{L$_{\odot}$}}
\newcommand{\jyb}{\mbox{Jy~beam$^{-1}$}}
\newcommand{\mjyb}{\mbox{mJy~beam$^{-1}$}}

\begin{document}

\newcommand{\gsim}{\raisebox{-.4ex}{$\stackrel{>}{\scriptstyle \sim}$}}
\newcommand{\lsim}{\raisebox{-.4ex}{$\stackrel{<}{\scriptstyle \sim}$}}
\newcommand{\psim}{\raisebox{-.4ex}{$\stackrel{\propto}{\scriptstyle \sim}$}}
\newcommand{\doce}{\mbox{$^{12}$CO\ }}
\newcommand{\trece}{\mbox{$^{13}$CO\ }}
\newcommand{\jcc}{\mbox{$J$=5$-$4}}
\newcommand{\jtd}{\mbox{$J$=3$-$2}}
\newcommand{\jdu}{\mbox{$J$=2$-$1}}
\newcommand{\juc}{\mbox{$J$=1$-$0}}
\newcommand{\mloss}{\mbox{$\dot{M}$}}
\newcommand{\my}{\mbox{$M_{\odot}$~yr$^{-1}$}}
\newcommand{\ls}{\mbox{$L_{\odot}$}}
\newcommand{\ms}{\mbox{$M_{\odot}$}}
\newcommand{\mm}{\mbox{$\mu$m}}
\def\arcdeg{\hbox{$^\circ$}}
\newcommand{\secp}{\mbox{\rlap{.}$''$}}

\title{ALMA high spatial resolution observations of the dense molecular region of NGC~6302}

\author{M. Santander-Garc\'\i a\inst{1,2}
\and V. Bujarrabal\inst{3}
\and J. Alcolea\inst{2}
\and A. Castro-Carrizo\inst{4}
\and C. S\'anchez Contreras\inst{5} 
\and G. Quintana-Lacaci\inst{1}
\and R.~L.~M. Corradi\inst{6}
\and R. Neri\inst{4}
}

\institute{
Instituto de Ciencia de Materiales de Madrid (CSIC), E-28049, Madrid, Spain \\ email: {\tt msantander@icmm.csic.es}
\and
Observatorio Astron\'omico Nacional (IGN), C/ Alfonso XII 3, E-28014, Madrid, Spain
\and
Observatorio Astron\'omico Nacional (IGN), Ap.\ de Correos 112, E-28803, Alcal\'a de Henares, Madrid, Spain
\and
Institut de Radioastronomie Millim\'etrique, 38406, Saint Martin d'H\`eres, France
\and
Centro de Astrobiolog\'\i a, CSIC-INTA, E-28692 Villanueva de la Cañada, Madrid, Spain.
\and
Instituto de Astrof\'\i sica de Canarias, E-38200 La Laguna, Tenerife, Spain
}



\abstract
 {The mechanism behind the shaping of bipolar planetary nebulae is still poorly understood. It is becoming increasingly clear that the main agents must operate at their innermost regions, where a significant equatorial density enhancement should be present and related to the collimation of light and jet launching from the central star preferentially towards the polar directions. Most of the material in this equatorial condensation must be lost during the asymptotic giant branch as stellar wind and later released from the surface of dust grains to the gas phase in molecular form. Accurately tracing the molecule-rich regions of these objects can give valuable insight into the ejection mechanisms themselves.} 
{We investigate the physical conditions, structure and velocity field of the dense molecular region of the planetary nebula NGC~6302 by means of ALMA band 7 interferometric maps.}
{The high spatial resolution of the \doce and \trece \jtd\ ALMA data allows for an analysis of the geometry of the ejecta in unprecedented detail. We built a spatio-kinematical model of the molecular region with the software {{\tt SHAPE}} and performed detailed non-LTE calculations of excitation and radiative transfer with the {{\tt shapemol}} plug-in.}
{We find that the molecular region consists of a massive ring out of which a system of fragments of lobe walls emerge and enclose the base of the lobes visible in the optical. The general properties of this region are in agreement with previous works, although the much greater spatial resolution of the data allows for a very detailed description. We confirm that the mass of the molecular region is 0.1 \msun. Additionally, we report a previously undetected component at the nebular equator, an inner, younger ring inclined $\sim$60$^\circ$ with respect to the main ring, showing a characteristic radius of 7.5$\times$10$^{16}$~cm, a mass of 2.7$\times$10$^{-3}$ \msun, and a counterpart in optical images of the nebula. This inner ring has the same kinematical age as the northwest optical lobes, implying it was ejected approximately at the same time, hundreds of years after the ejection of the bulk of the molecular ring-like region. We discuss a sequence of events leading to the formation of the molecular and optical nebulae, and briefly speculate on the origin of this intriguing inner ring.}
{}

\keywords{Interstellar medium: kinematics and dynamics -- planetary nebulae: general -- planetary nebulae: individual: NGC~6302, Physical data and processes: radiative transfer
}

\titlerunning{ALMA high spatial resolution observations of the dense molecular region of NGC~6302}

\maketitle
%

\section{Introduction}

The question of how low- and intermediate-mass stars depart from spherical symmetry during the latest stages of stellar evolution, producing gaseous nebulae that are mostly bipolar, is one of the most intriguing, open topics in the field (e.g. \citealp{balick02}). Tracking the mass ejecta of planetary nebulae (PNe) by means of molecular emission lines can provide crucial clues for better understanding the shaping of these objects, especially in the innermost regions around the central stars. As an asymptotic giant branch (AGB) star evolves towards the post-AGB stage, its effective temperature increases and a photodissociation region (PDR) is developed around it as the ultraviolet radiation progressively erodes the innermost remnants of the expanding AGB wind. Although the inner region of some post-AGB and young PNe has been mapped down to $\sim$200 AU, this innermost gap devoid of molecules has never been resolved (e.g. the Red Rectangle; \citealp{bujarrabal13}).

In this work, we present high spatial resolution Atacama Large Millimeter/submillimeter Array (ALMA) observations of NGC~6302 (PN G349.5+01.0), tracing emission from the \doce and \trece\ \jtd\ transitions. NGC~6302 is a large, butterfly-like young PNe located 1.17~kpc away (\citealp{meaburn08}) whose central star left the AGB 2,100 years ago according to photoionisation models, and has an effective temperature of 220,000 K and a luminosity of 14,300 \lsun\ (\citealp{wright11}). It consists of a system of high-velocity bipolar outflows in optical images with a kinematical age of 2,200 years, and a massive (0.1 \msun), seemingly fragmented and complex molecular disk in the equator of the optical nebula showing expanding kinematics (\citealp{peretto07,santander15a,dinhvtrung08}). These authors analysed the physical conditions (i.e. densities, temperatures, and abundances) and the velocity field of the molecular region, and determined that the disk had been ejected during 4,500-5,000 years, ending $\sim$2,900 years ago, but were unable to provide much information on the geometry of the ejecta because of the limited spatial resolution of their data. The spatial resolution achieved by the ALMA observations of this object, together with previous single-dish data from James Clerk Maxwell Telescope (JCMT) and the Heterodyne Instrument for the Far Infrared (HIFI) on board the Herschel Space Observatory, allows us to build a {\tt SHAPE+shapemol} model to determine its physical conditions, describe the velocity field, and investigate the geometry of the ejecta in unprecedented detail (\citealp{steffen11,santander15a}).

The structure of the paper is as follows. The ALMA observations are described in section 2; section 3 deals with the data description and the iterative process followed for modelling the molecular ejecta of NGC~6302. The results of the best-fit model are detailed in section 4, discussed in section 5, and summarised in section 6.

\section{Observations}

We observed NGC~6302 with the ALMA interferometer in June 14, 2014 (programme ID 2013.1.00331.S) with 35 antennas in an extended configuration and dual polarisation. We mapped the rotational transitions \doce and \trece \jtd\ in band 7, 345.796 and 330.588 GHz (see the {\tt Cologne Database for Molecular Spectroscopy} by \citealp{muller01,muller05} for the line identification used in this work), respectively, in two of the spectral windows, while the other two picked up emission from the nearby continuum. Titan was observed as flux calibrator, while the phase and amplitude of the observations were calibrated with J1709-3525. J1427-4206 was used for the bandpass calibration. We obtained a total of 20 min of correlation time on NGC~6302. The backends were set to achieve a spectral resolution of 0.2 \kms\ in the lines, with a total observed frequency span between 345.736 and 345.854 GHz, and between 330.529 and 330.647 GHz, respectively, and a spectral resolution of 1.7 \kms\ in the two spectral windows used for continuum subtraction of the lines, with observed ranges 343.413--345.288 (upper side band continuum) and 330.813--332.688 GHz (lower side band continuum).

We improved the delivered {\tt CASA}\footnote{Common Astronomy Software Applications, developed by NRAO, ESO, NAOJ, CSIRO/ATNF, and ASTRON.}-calibrated data by recalibrating them according to our specific needs. We performed the cleaning in {\tt GILDAS}\footnote{GILDAS is a collection of state-of-the-art softwares oriented towards (sub-)millimeter radioastronomical applications; http://www.iram.fr/IRAMFR/GILDAS} with the aid of support polygons enclosing the extended emission. Once we confirmed there was no emission arising from any central, low-velocity, very small-scale structure at the best spatial resolution achievable by uniform weighting, we averaged the initial velocity channels, obtaining velocity resolutions of 0.42 \kms\ in the lines, and performed image deconvolution by robust weighting (with a value of 0.5). The resulting brightness rms is $\sim$6 \mjyb\ in maps from both \doce and \trece \jtd\ emissions, and $\sim$1.7~\mjyb\ in the continuum. The synthetic beam of the \doce \jtd\ map is a 2D Gaussian with a half power beam width (HPBW) of 0\farcss 38$\times$0\farcss 33, with its major axis oriented along a PA of 87$^\circ$. The synthetic beam of the \trece\ \jtd\ map, on the other hand, has a HPBW of 0\farcss 43$\times$0\farcss 36, with its major axis oriented along a PA of 97$^\circ$. The conversion factors from flux units to main-beam brightness temperature (T$_\mathrm{mb}$) units are 79.5~K and 72.5~K per Jy~beam$^{-1}$ in the \doce\ and \trece\ \jtd\ maps, respectively.

Velocity data in this work are referred to the local standard of rest. Conversion to \vhel\ for the coordinates of NGC~6302 is given by  \vhel$ = $\vlsr$ - 7.45$ \kms.

\subsection{Flux loss}

\begin{figure}[!]
\center
\resizebox{8.5cm}{!}{\includegraphics{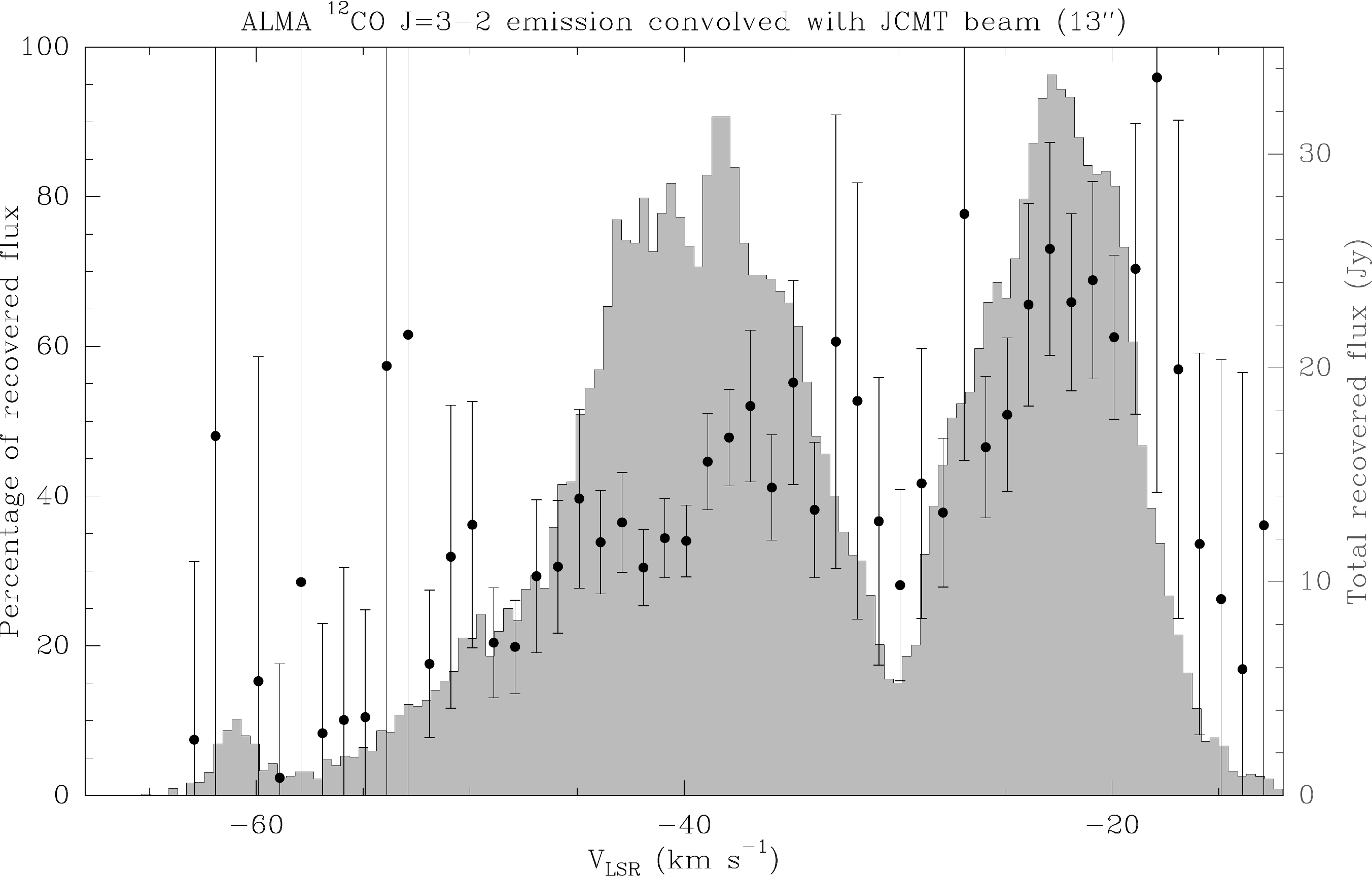}}
\caption{Percentage of the recovered flux in the ALMA observations of \doce \jtd\ when compared to JCMT observations. The points show the percentage of recovered flux across channels (left vertical axis), with error bars including uncertainties in the calibration of both instruments. The filled, grey histogram corresponds to the total recovered flux profile of NGC~6302 in the \doce \jtd\ ALMA maps convolved with the HPBW of JCMT at 345 GHz and at the position observed by \citeauthor{peretto07} (right vertical axis). The emission around \vlsr=-62 \kms, which is not present in the \trece\ \jtd\ map, is actually contamination from NS (see Fig.~\ref{Fothermol}).}
\label{Flost12}
\end{figure}

\begin{figure}[!]
\center
\resizebox{8.5cm}{!}{\includegraphics{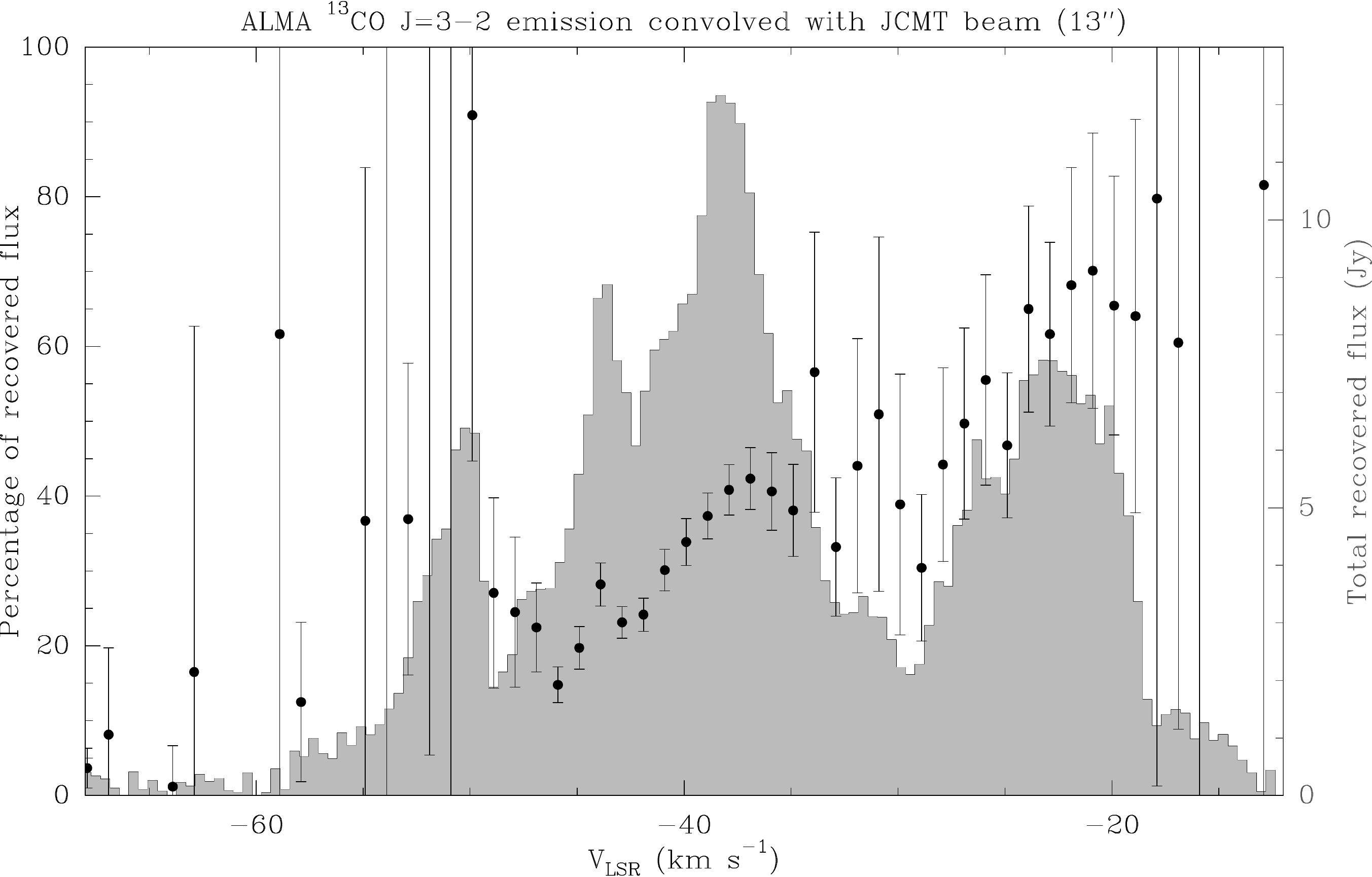}}
\caption{Percentage of the recovered flux in the ALMA observations of \trece \jtd\ when compared to JCMT observations. The points show the percentage of recovered flux across channels (left vertical axis), with error bars including uncertainties in the calibration of both instruments. The filled, grey histogram corresponds to the total recovered flux profile of NGC~6302 in the \doce \jtd\ ALMA maps convolved with the HPBW of JCMT at 330 GHz and at the position observed by \citeauthor{peretto07} (right vertical axis).}
\label{Flost13}
\end{figure}

The phase centre for our ALMA observations were the same as in SMA observations by \cite{dinhvtrung08}, that is R.A. 17$^h$13$^m$44.21$^s$, Dec. -37$^{\circ}$06$^\prime$15.941$^{\prime\prime}$. This position is offset by 5\farcss 4 from the centre of the molecular rings described in Sect. 3. Also, a significant flux is found to be filtered out in the interferometric data due to the missing short spacings. We estimated the recovered flux by comparison of the spectral profiles in JCMT observations by \cite{peretto07} with the profiles of the \doce\ and \trece\ \jtd\ maps convolved with the HPBW of the JCMT at the corresponding frequencies, and taken at the position of the JCMT observations. 

The recovered flux at each channel is shown in Figs.~\ref{Flost12} and ~\ref{Flost13}, along with error bars. The amount of recovered flux is highly dependent on the channel, being $\sim$70\% in the redwards regions, between 40 and 50\% in the blueward regions, and $\sim$30\% near the systemic velocity, that is, in the regions close to the plane of the sky, where any large-scale smooth structure tends to be completely washed out by interferometric filtering.

\begin{figure*}[!]
\center
\resizebox{18cm}{!}{\includegraphics{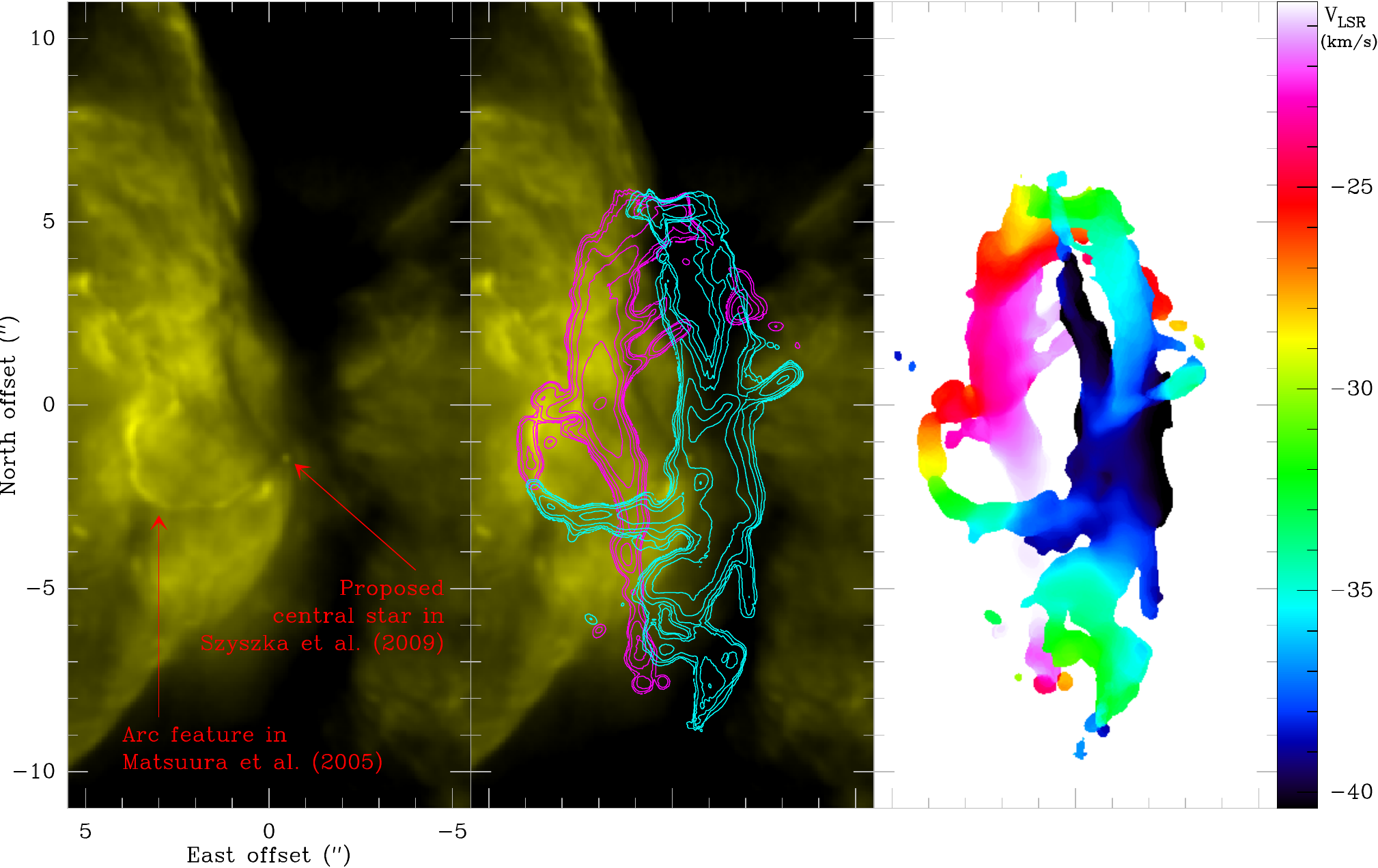}}
\caption{{\bf Left panel:} HST/WFPC2 F673N image of the central region of NGC~6302. The intensity levels have been chosen to highlight the filament (or `arc', as depicted by \citealp{matsuura05}), which seems to be the ionised inner rim of the inner ring reported in this work. {\bf Middle panel:} Integrated \doce \jtd\ emission in the 20 \kms\ around the systemic velocity of NGC~6302, with a threshold of 0.1 \jyb. Cyan contours correspond to the emission from $V_\mathrm{LSR}$ -40.4 to -30.4 \kms and magenta contours to the emission from -30.4 to -20.4 \kms. The first contour corresponds to a level of 1 Jy~\kms, with a multiplicative step of $\sqrt{3}=1.732$ between adjacent contours. Both rings are apparent in the figure, the main ring running almost vertically, the other inclined, and its major axis running almost along the northwest to southeast direction. Some contours inside the main ring denote a brightness decrease towards holes with no emission (shown in white in the right panel). {\bf Right panel:} Intensity-weighted mean radial velocities in the same intervals as in the middle panel, revealing an expanding kinematics for both structures. The colour-velocity correspondence is shown to the right of the panel. The systemic velocity is $V_\mathrm{LSR}$=-30.4 \kms.}
\label{Fmoments}
\end{figure*}

\section{Observational results and data modelling}

The ALMA observations show emission lines from $^{12}$CO, $^{13}$CO, NS, and SO. The latter two are considerably fainter than those of CO and have not been included in our modelling. A brief description of the emission from these lines is provided in Appendix A.

The \doce\ and \trece\ \jtd\ observations are largely compatible with previous \jdu\ maps by \cite{dinhvtrung08}, except for the larger amount of flux lost in present observations with ALMA. The much higher spatial resolution of these maps nevertheless allow us to delve deeper into the geometry of the ejecta. The \doce\ \jtd\ observations are shown in Figs.~\ref{Fmoments}, \ref{F12obs}, and \ref{F12obsring}, while Figs.~\ref{Fsketch}, \ref{F12mod}, and \ref{F12modring} show a sketch of the model and our best-fit to these maps; see Sect. 4. Emission from NS, SO, \trece\ \jtd, and a model for the latter, are shown in  Appendices A and B (Figs.~\ref{FmapSO}, \ref{FmapNS}, \ref{F13obs}, and \ref{F13mod}). The observations show a large and broken ring or torus, out of which curved structures emerge, seemingly running parallel to the base of the lobes visible in the optical. In order to test this, we used an optical image in the F673N filter from the Hubble Space Telescope (HST) archive (cycle 15, Director's Discretionary Program 11076, PI Noll, K.) and used it for comparison with the CO data, once astrometrically corrected to an accuracy of 0\farcss 1 using field stars registered with the fourth US Naval Observatory CCD Astrograph Catalog (UCAC4, \citealp{zacharias12}).

When computing the moments of the emission from $V_\mathrm{LSR}$=-63 to -12 \kms\ and detection thresholds 15 and 10 \mjyb\ for \doce\ and \trece respectively, we find that the integrated emission seems to follow the warped, extended disk visible in optical images, although the amount of warping is negligible in the region holding the bulk of CO emission.

The ALMA maps have allowed us to detect and analyse a new structure expanding inside the cavity left by the expanding, main molecular ring. This structure consists of another, smaller and thinner ring of material which, when the data are overlaid on HST images, follows the feature labelled as `arc' by \cite{matsuura05} in their Fig.~8 (see Fig.~\ref{Fmoments} in this work). Both the main and inner rings are made apparent when the integrated emission and the intensity-weighted mean velocities of NGC~6302 are overlaid on a F673N HST image in Fig.~\ref{Fmoments}. This new, inner ring is further analysed and discussed in the present work. 

Our figures are focused on the centre of this inner ring, located at R.A. 17$^h$13$^m$44.455$^s$, Dec. -37$^{\circ}$06$^\prime$11.41$^{\prime\prime}$ (J2000). This position lies approximately at the centre of the 3.6~cm emission studied by \cite{gomez93} with the VLA, while  offset by 1\farcss 7 approximately to the north of the position of the unseen central star as proposed by \cite{szyszka09}.

The model of the molecule-rich region surrounding the central star of NGC~6302 presented in this work is based on previous results, mainly by \cite{dinhvtrung08} and \cite{santander15a}, and uses the morphology, velocity field, densities, temperatures, and abundances found by the latter as a starting point. It is important to stress that any assumption on the geometry, orientation, and velocity field arise from direct interpretation of the data themselves.

Previous models focused on the physical conditions of the molecule-rich gas of this object rather than on providing an accurate description of its geometry because of the lack of spatial information attainable by single-dish observations or by maps with arcsec-wide synthetic beams. The high spatial and velocity resolution achieved in ALMA \doce and \trece \jtd\ maps allow us to derive the spatial distribution of gas in great detail. Unfortunately we did not have short-spacing data to image and analyse the smooth extended structures. This is especially troublesome near the systemic velocity \vlsr = -30.4 \kms\ (see below, Figs.~\ref{Flost12} and \ref{Flost13}), that is, in structures moving in the plane of the sky, where only a $\sim$30\% of the flux is recovered. No certain conclusions can be drawn from those regions, so any potential structure expanding outwards along the plane of the sky has been deliberately excluded from the model. This flux-loss caveat, however, has a positive counterpart, as interferometric filtering tends to filter out the most extended emission structures, thus allowing us to see smaller scale structures that would remain otherwise undetected. This might be the case for the faintest part of the inner ring in the northwest region of the molecular region of NGC~6302.

We initially investigated the general shape of the molecular region by reconstructing its 3D geometry from the ALMA maps by assuming a Hubble-like flow with different slopes (kinematic ages). The $x$ and $y$ position of every point in every channel above an emission threshold (defined at a 3$\sigma$ level) was registered in a text file along with a $z$ position computed from the specific velocity law assumed. The text file with the positions was then loaded into {\tt SHAPE} and visually inspected. The presence of a broken ring with emerging fragments of lobes and of an inner ring were inferred, and their approximate kinematical ages that were derived by changing the velocity law until the corresponding ring (the main or the inner one) showed a circular shape in the reconstruction. We then used those kinematical ages as starting points for the modelling of each structure. We also investigated alternative reconstructions using constant velocity laws, but the structures generated in that manner showed very irregular shapes without any degree of axial symmetry, so we discarded such a velocity law except for the inner regions of the main ring (see Section 4).

In order for the model to overcome flux loss and be coherent with all the data, we fitted the geometry and velocity patterns of the structures visible in the ALMA \doce and \trece maps, and then fitted the intensities of previously existing data from single-dish telescopes, which are unaffected by flux loss. These include Herschel/HIFI high-$J$ transitions observed by \cite{bujarrabal11} and modelled by \cite{santander15a}, and the JCMT \doce\ and \trece\ \jtd\ observations by \cite{peretto07}.  However, this strategy tends to underestimate the intensity levels of synthetic single-dish profiles at certain velocities, since their corresponding observations do not suffer from flux loss. We therefore opted to model the structures reaching a compromise between the somewhat overestimated maps and the slightly underestimated single-dish profiles.

We used {\tt SHAPE}+{\tt shapemol} to produce a detailed 3D model and generate high-resolution synthetic spectral profiles and maps to be compared with observational data. {\tt SHAPE} is an interactive 3D software for modelling complex gaseous nebulae. The geometry of the nebula and its density, temperature, and velocity distributions are generated in an interactive way using 3D mesh structures via a user-friendly interface. The code is able to generate synthetic images, position-velocity diagrams, spectral profiles, and maps for direct comparison with observations. {\tt shapemol} works as a plugin for {\tt SHAPE}, enabling full radiative transfer in \doce\ and \trece\ lines by using a set of pre-generated tables computed under the assumption of the large velocity gradient (LVG) approximation (see e.g. \citealt{castor70}) to characterise the absorption and emission coefficients of every cell in the 3D grid, for the abundances, species and desired transitions specified by the user. See \cite{santander15a} for a detailed description of the code, its capabilities and limitations.

Given the complexity of the model, with a potentially infinite number of degrees of freedom, the fitting procedure is performed by visual inspection on a series of iterations, modifying one parameter at a time. When a reasonable fit to all data is found, the vicinity of the parameter space is carefully explored for better solutions. In order to avoid reaching a parameter space local minimum, several series of modelling iterations are performed, each starting from a different set of parameters over a large range. Uncertainties are estimated by bounding the range of values of the characteristic sizes, densities, and temperatures that lead to acceptable predictions. Uncertainties on the velocities are derived by eye, by varying the different velocity fields and the systemic velocity until the model no longer provides a good fit. 

\section{Modelling results}

\begin{figure*}[!]
\center
\resizebox{18cm}{!}{\includegraphics{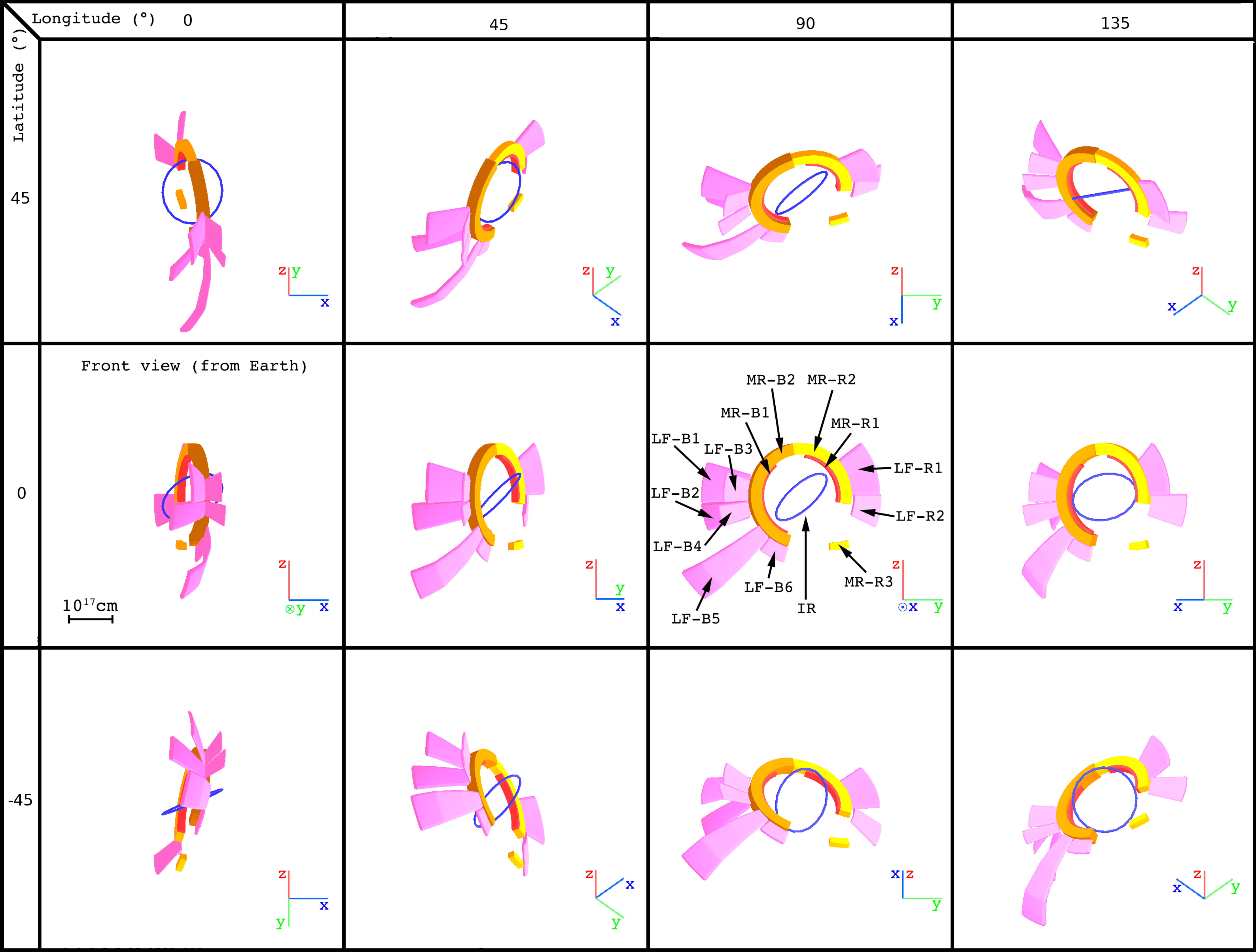}}
\caption{Model sketch of the molecular region of NGC~6302 as seen from different orientations. The reference system is centred on the model, with the $x$-axis towards west; the $y$-axis along the line of sight, away from Earth; and the $z$-axis towards the north. Each inset shows a different perspective of the model with the camera orbiting around it, positioned at different latitudes and longitudes as conventionally defined in spherical coordinates. The scale is indicated in the `0,0' inset, which corresponds to the view from Earth and is therefore labelled as `Front view'. The projection of the reference axes is shown in each inset to aid visualisation. The different structures are indicated by labels and arrows with the following syntax: IR, inner ring; MR, main Ring; LF, lobe fragment'. B and R stand for blueshifted and redshifted, respectively.}
\label{Fsketch}
\end{figure*}

\begin{figure*}[!]
\center
\resizebox{18cm}{!}{\includegraphics{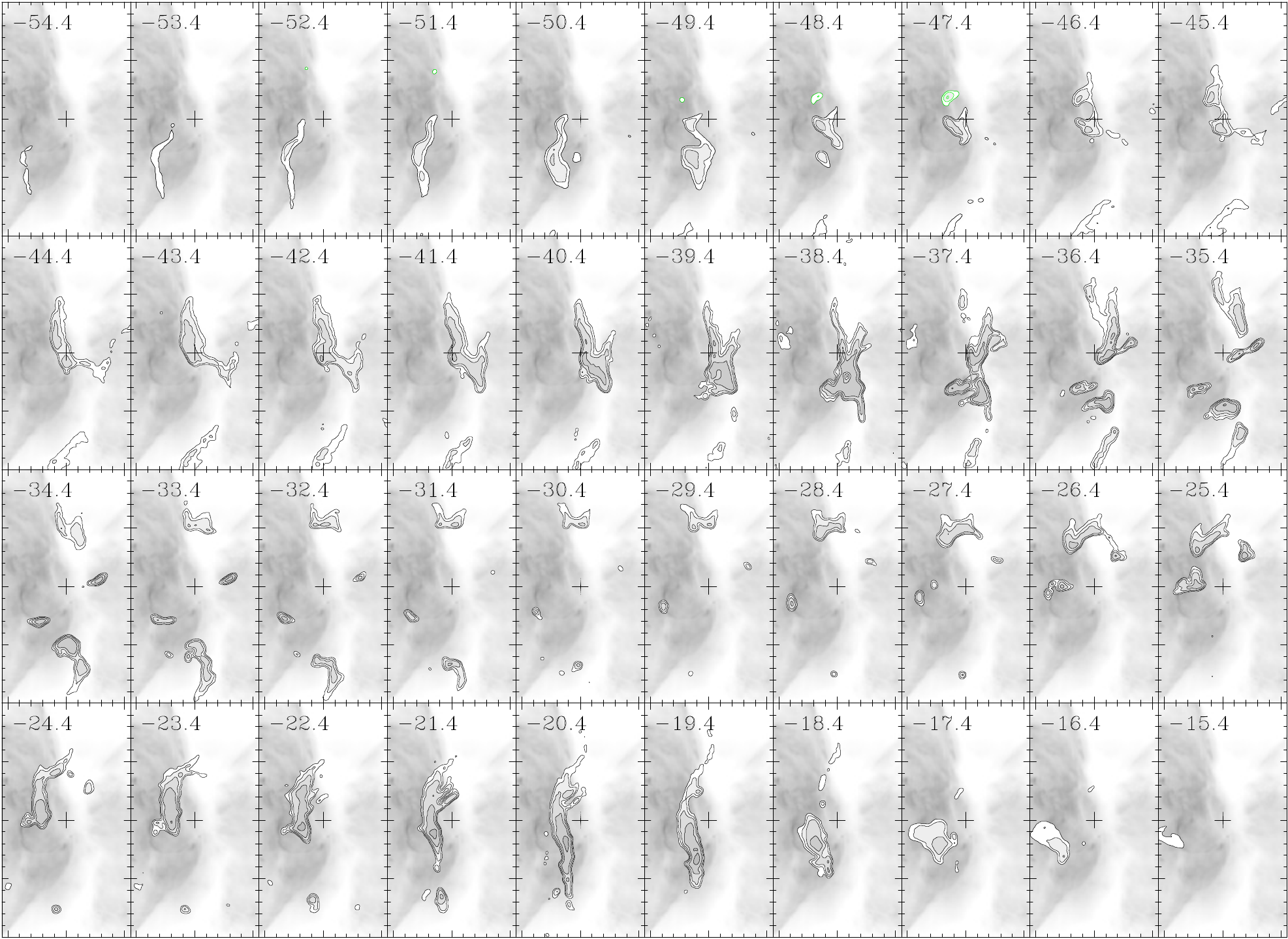}}
\caption{ALMA \doce \jtd\ emission map of NGC~6302, binned to 1 \kms. The field of view is 11$''\times20''$. \vlsr\ velocities in \kms\ are indicated in each channel. The intensity scale of the contours is logarithmic with the first contour at 40 \mjyb, and a step of 3 dB (i.e. a factor of 10$^{0.3}=2$) between adjacent contours. The deepest negative contour, only visible in the 38.0 \kms channel, has a level of -100 \mjyb. Contamination by NS emission (see Appendix A) is highlighted in green wherever it can be spatially resolved from \doce\ emission. An HST F673N image of the central region of NGC~6302 is shown in the background for comparison purposes.}
\label{F12obs}
\end{figure*}

\begin{figure*}[!]
\center
\resizebox{18cm}{!}{\includegraphics{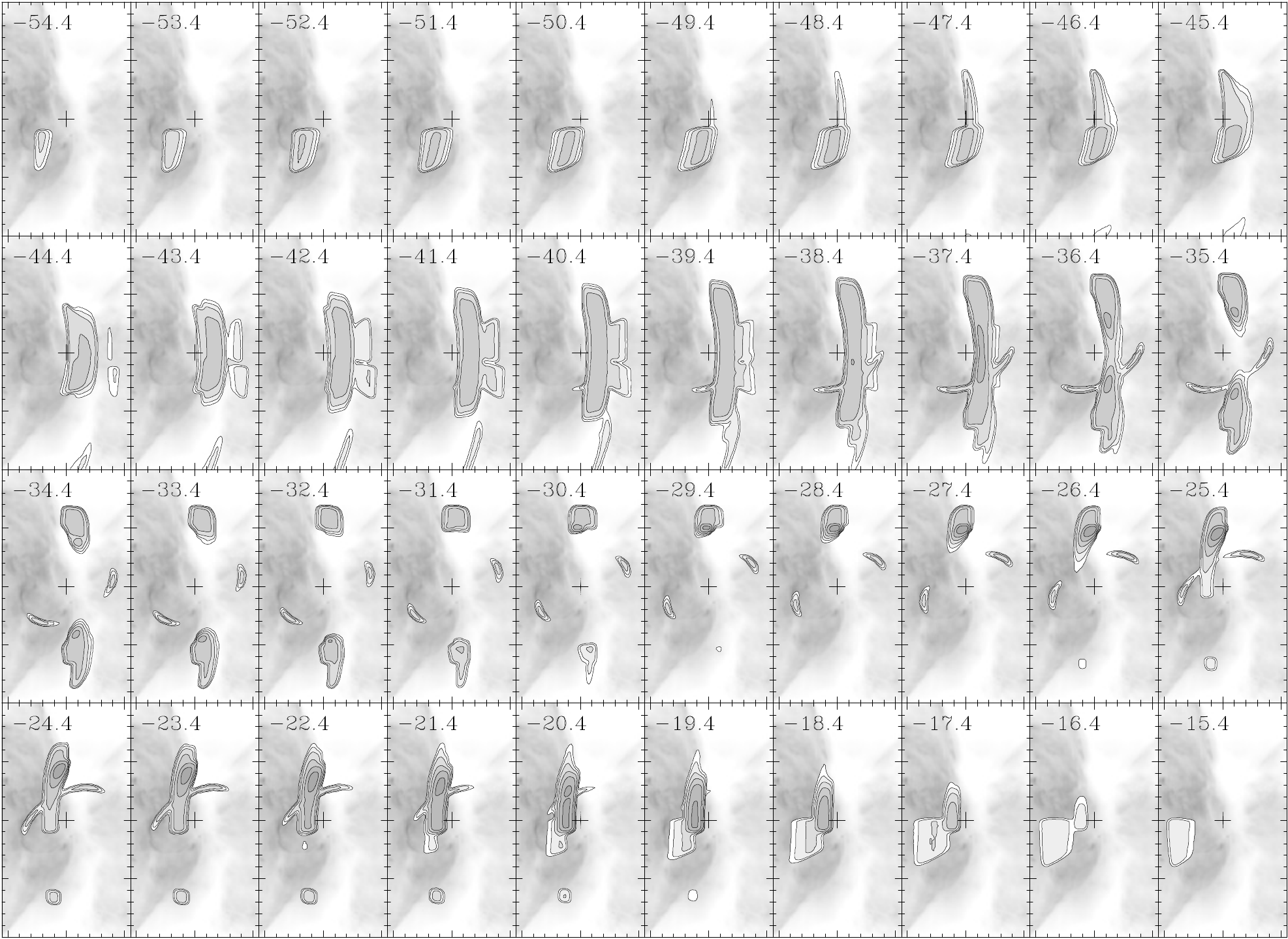}}
\caption{ALMA \doce \jtd\ model of the molecular region of NGC~6302. The field of view is 11$''\times20''$. \vlsr\ velocities in \kms\  are indicated in each channel. The intensity scale of the contours is logarithmic, with first contour at 40 \mjyb, and a step of 3 dB (i.e. a factor of 10$^{0.3}\simeq2$) between adjacent contours. The differences in thickness and intensity of the model when compared to the observations mainly arise from the considerable flux loss affecting the ALMA observations (see text), which are particularly noticeable near the systemic velocities. Also, the logarithmic scale gives a false, too pessimistic impression of the brightness overestimate of the synthetic maps, which mainly affects the first, very low contours. An HST F673N image of the central region of NGC~6302 is shown in the background for comparison purposes.}
\label{F12mod}
\end{figure*}

\begin{figure*}[!]
\center
\resizebox{18cm}{!}{\includegraphics{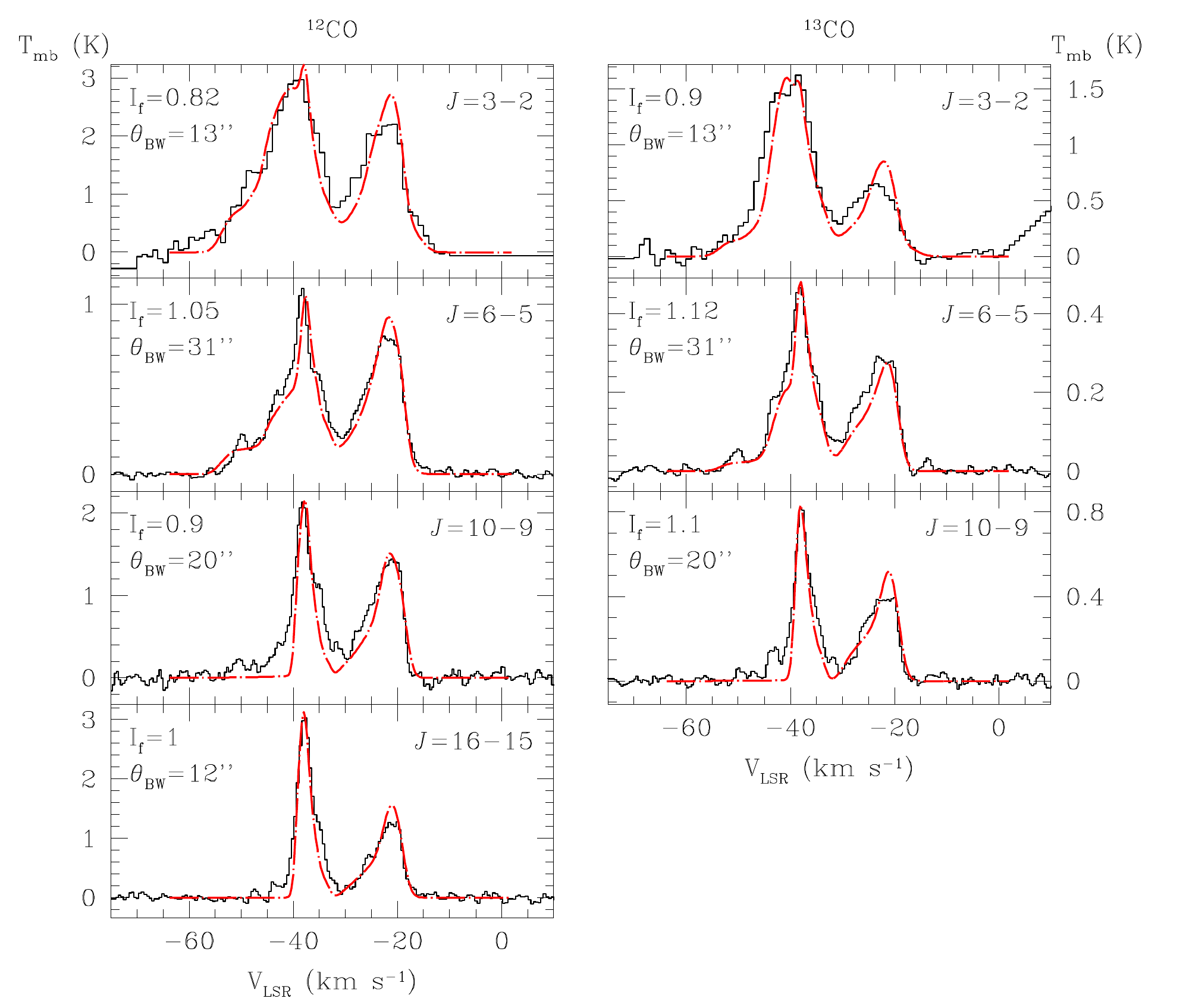}}
\caption{Resulting synthetic spectra (red, dot-dashed line) and observations (black histogram) for the \doce\ and \trece\ transitions detected in NGC~6302 with JCMT ($J$=3-2) and HERSCHEL/HIFI ($J$=6-5, 10-9, and 16-15). The parameter $\theta_{BW}$ stands for the HPBW of the telescope beam at the given frequency, while $I_\mathrm{f}$ refers to the intensity factor applied to the model to account for the uncertainties in the radiative transfer solving of the model and in the calibration of the observations, which enable a good fit to each spectrum.}
\label{Fsingle}
\end{figure*}

\begin{figure*}[!]
\center
\resizebox{12.5cm}{!}{\includegraphics{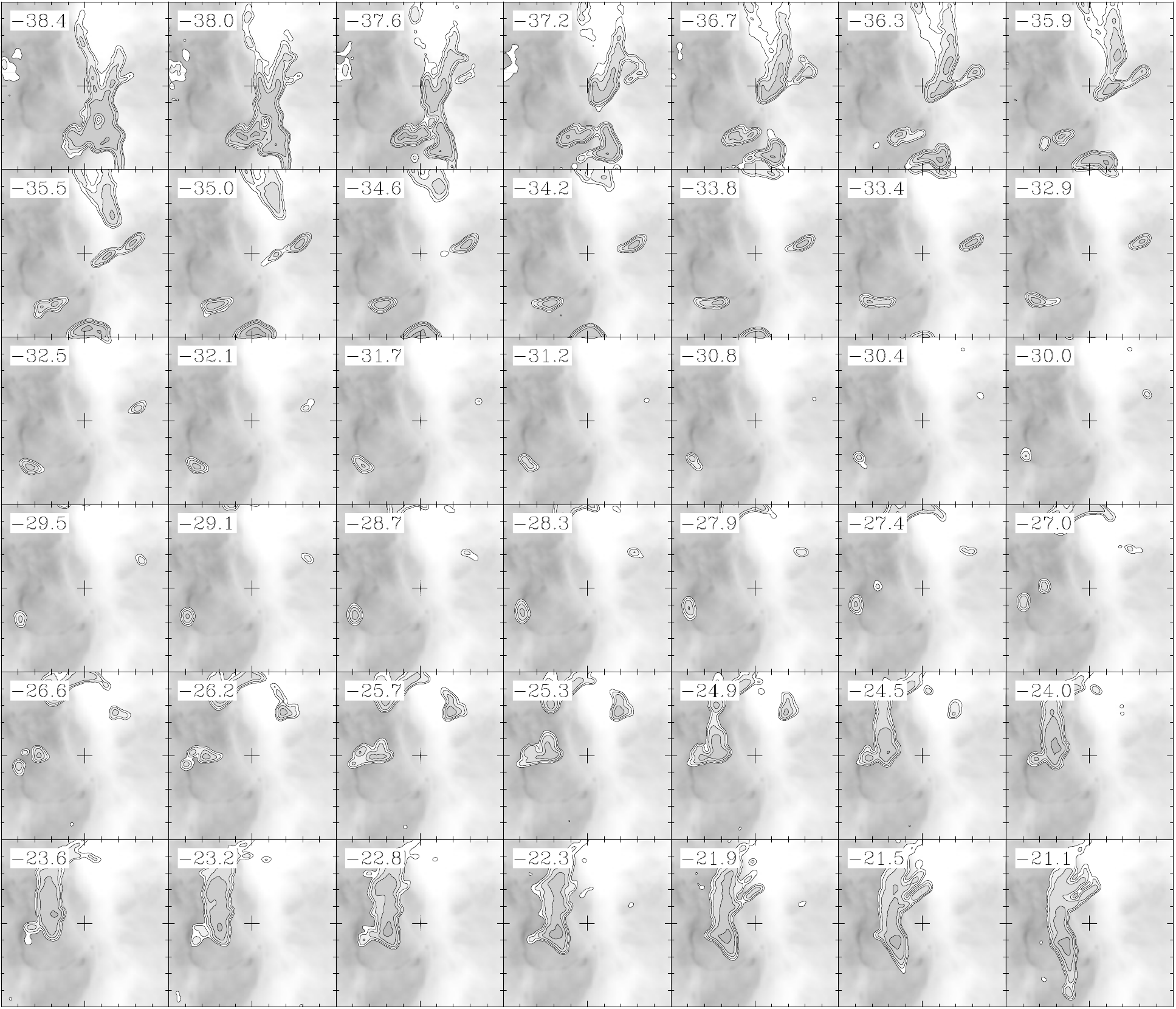}}
\caption{Close-up of the emission from the inner ring in the central velocities of the  \doce \jtd\ ALMA map (see Fig.~\ref{F12obs}). The field of view is the central 10$''\times10''$. The intensity scale of the contours is logarithmic, with first contour at 40 \mjyb, and a step of 3 dB (i.e. a factor of 10$^{0.3}\simeq2$) between adjacent contours. The background is an HST F673N image for comparison purposes.}
\label{F12obsring}
\end{figure*}

\begin{figure*}[!]
\center
\resizebox{12.5cm}{!}{\includegraphics{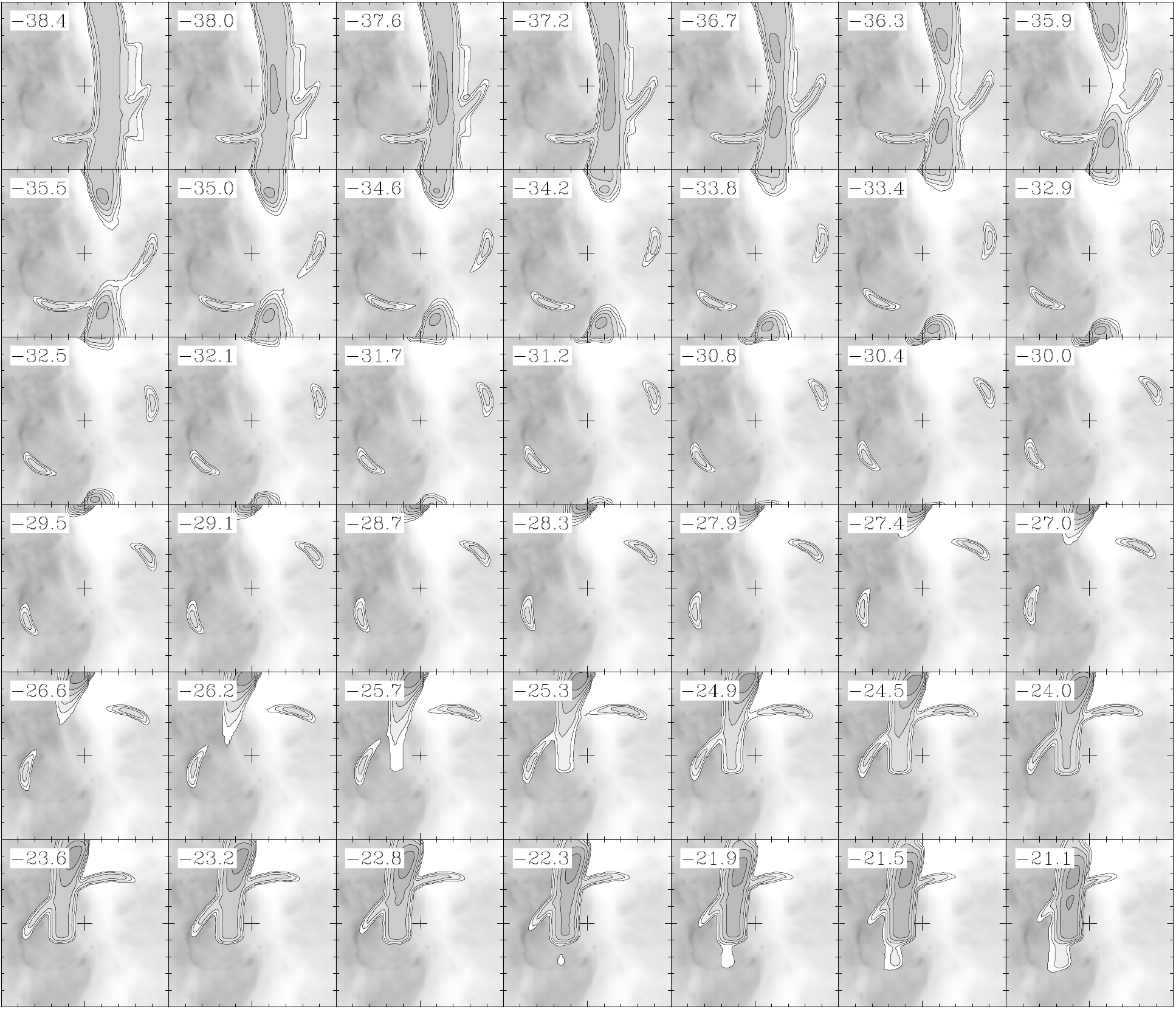}}
\caption{Close-up of the emission from the inner ring in the \doce \jtd\ model. The field of view is 10$''\times10''$. The intensity scale of the contours is logarithmic, with first contour at 40 \mjyb, and a step of 3 dB (i.e. a factor of 10$^{0.3}\simeq2$) between adjacent contours. The background is an HST F673N image for comparison purposes.}
\label{F12modring}
\end{figure*}

Sketches of the best-fit model from different points of view are shown in Fig.~\ref{Fsketch}. It consists of a broken ring of material whose symmetry axis is inclined 75\arcdeg\ to the line of sight. The resulting ellipse in the plane of the sky has its major axis oriented along the north-south direction (P.A.=0\arcdeg). The large-scale warping of this structure, visible in optical images, has not been considered in this work as it is negligible in the bulk of the CO emission revealed by the ALMA maps (see Fig.~\ref{Fmoments}). This main ring is split into an inner, hot, and denser region, at a distance between 8.8 and 9.5$\times$10$^{16}$~cm from the star, and an outer, cool, and more tenuous region, at a distance from the star between 9.5 and 12.1$\times$10$^{16}$~cm. A system of lobe fragments emerges from the outskirts of this main ring, running outside of and approximately parallel to the base of the lobes seen in the optical, resembling the sepals of a flower as they enclose the base of the petals. This is made apparent when displaying the \doce and \trece map contours over the F673N HST image. These lobe fragments reach out to $\sim$3$\times$10$^{17}$ cm from the central star and are broken into holes and clumps of gas. A new feature inside the main ring, which is absent in previous modellings, is found to be well characterised by a ring or torus with inner and outer radii of 7.2 and 7.7$\times$10$^{16}$~cm, respectively; this feature is at a very different orientation to the main rings, that is, inclined 52\arcdeg\ to the line of sight and with the major axis of the projected ellipse along P.A.=119\arcdeg\ (see Fig.~\ref{Fsketch}). The actual geometry of this structure, however, is more complex than that of a simple ring, since the knots in the northwest part seem to be slightly closer to the star than their southeastern counterparts (see the best-fit to the maps in Figs.~\ref{F12obsring} and \ref{F12modring}). This could result from a slightly spiral pattern or from the action of shocks on the ring, although modelling this feature as a more complex structure with the data at hand is beyond the scope of this work.

All of these structures are directly identifiable in the emission of the ALMA \doce\ and \trece\ maps, as the main ring and emerging lobe fragments are associated with the broken torus identified and studied by \cite{dinhvtrung08} and \cite{santander15a}. Consideration of the high-$J$ CO rotational lines from Herschel/HIFI leads to the identification of the corresponding emission from the main structures except for the inner ring, and provides additional information about the stratified distribution of the temperature and density of the molecular gas.

Each of the structures of the model is characterised by a constant value of the temperature, and a density that drops with the inverse of the distance to the central star (except in the inner ring and the inner rim of the main ring, where densities are constant). The velocity pattern of every structure is radial, expanding outwards from the central star. Every structure shows a velocity pattern that is characterised by a linear dependence on the distance to the star (a Hubble-like flow), except for the thin, inner sections of the main ring, which show a constant value of the velocity that is larger than that of the outer section.

The systemic velocity of the nebula deduced from our modelling is \vlsr = -30.4 \kms, in agreement with the -30 \kms\ found by \cite{peretto07}. We have set the \doce and \trece abundances to single values for the whole nebula. We estimate the abundance ratio [\doce]/[\trece] ratio to be 8, a value that is somehow low but not unexpected in post-AGB nebulae, particularly in O-rich stars (see discussion in \citealp{bujarrabal13}), and even in O-rich AGB stars (e.g.~\citealp{milam09}, \citealp{debeck10}). The best-fit \doce and \trece abundances are 1$\times$10$^{-4}$ and 1.25$\times$10$^{-5}$, respectively, which is a factor 1.5 and 2 lower than in previous modellings by \cite{santander15a}. The reason behind this discrepancy is partly due to the current model of the main ring and emerging fragments of lobes. This model is slightly larger as a result of the more accurate measurement of the surface brightness distribution, and partly because of the densities and temperatures, which have to be similarly high to account for the intensity of the high-$J$ transitions. These circumstances necessarily lead to the determination of a lower value of the \trece abundance; the \doce abundance is less critical owing to its larger optical thickness.

Results are summarised in Table 1, along with the molecular mass and computed kinematic age of the different structures. The geometry of the system of lobe fragments is too complex to be coded into a single table and is therefore not included here, although a file provided in electronic form with the model of the dense molecular region of NGC~6302 in {\tt SHAPE} format is available as on-line supporting data. 

\subsection{Model uncertainties and reliability of the fit}

The best-fit model provides a reasonable compromise between the fit to the ALMA maps (see Figs.~\ref{F12obs} and \ref{F12mod} for the \doce\ \jtd\ emission), and to the low- and high-$J$ single-dish data (see Fig.~\ref{Fsingle}) of  JCMT and Herschel/HIFI. We applied different free intensity scale factors to the corresponding synthetic profiles to account for the characteristic 20\% uncertainty  of the calibration of this instrument (see \citealp{santander12}). These factors range from 0.85 to 1.12. The synthetic map of the \trece\ \jtd\ emission is very similar to its \doce\ counterpart, except for its intensity levels, and is thus shown in Appendix B along with the observations (see Figs.~\ref{F13obs} and \ref{F13mod}).

We chose to model the complex molecular region revealed by the observations with a relatively simple model characterised by simple velocity, density, and temperature laws, since a perfect fit to the data would require dealing with a parameter space that is too large to draw any solid conclusions. As such, the fit to the data is fair, with a number of caveats. In general, some components of the synthetic maps look considerably thicker than the emission in the ALMA observations. This discrepancy is partly due to the severe flux loss in the ALMA maps, which tend to highlight structural edges and fade away smooth, large regions, and partly to the intrinsic complexity of the molecular region itself, full of spurious channel to channel variations, clumps, holes and irregularities. Modelling these details would be out of the scope (and capability) of this work. 

In particular, we only modelled the most prominent lobe fragments, those which extend beyond the main ring. In light of the data, these lobe fragments seem to be part of the same structure as the main ring, which in turn shows a thin, X-shaped structure around \vlsr=-38.4 \kms \ that is not reproduced by the model. Part of this ring seems to be shaped as the base of a thin hourglass structure out of which the lobes would emerge in a continuous structure, as implied by the X-shaped feature seen in the \doce\ and \trece\ data from $\sim$-43 to $\sim$-37 \kms. We chose, however, to model the main ring as a thick cylindrical ring, since this hourglass structure appears very broken in the observational data and would therefore significantly increase the complexity of the model. The maps also show some condensation towards the south of the central star around \vlsr=-20 \kms; this region seems to be part of the main ring except for its velocity, which is larger than would be expected from the simple velocity pattern used to build the model. Therefore we chose to reproduce this feature as a distinct lobe fragment (BL-R2), for the sake of simplicity.

The inner ring also shows substructures that are not taken into account by our model. The northwest region of the ring is particularly clumpy and broken, showing very faint emission in several consecutive channels (mainly from \vlsr\ -32 to -30 \kms) and seemingly split into two thin structures around \vlsr = -21.5 \kms. Our approach has consisted of modelling this structure as a ring with a constant density, fitting its emission to the average intensity across channels in the observations.

In order to check for model consistency, we used {\tt GILDAS} to estimate the flux lost in simulated ALMA observations of the model in conditions as similar as possible to those of the actual data. The thin and clumpy structures visible in the observations are not recovered in this process, although the resulting flux loss in a series of tested channels across the velocity range in the synthetic observations is similar to that in the real data, thus providing consistency to the model. For instance, between 3 and 35\% of the flux is recovered at \vlsr = -40.4 \kms according to our simulations, while the percentage increases to between 82 and 98\% at \vlsr = -23.2 \kms. In channels near the systemic velocity of -30.4 \kms, between 80 and 100\% of the flux is recovered, as opposed to what occurs in observations, where almost 80\% of the flux is lost due to interferometric filtering of smooth structures. We consider this to be due to the limitations of the modelling, since we have excluded from the model any structures not detected in ALMA observations, and those structures probably include extended structures similar to the analysed lobe fragments running along the plane of the sky and, therefore, are prone to being filtered out.

Uncertainties in the density and temperature are provided in Table 1; these are computed by fixing all parameters but one, which is varied until the synthetic maps or single-dish profiles no longer provide a reasonable fit. Uncertainties in the velocity fields were derived by visual comparison, allowing the size and velocity of an structure to vary until its emission did not match the corresponding channels of the map anymore. 

Taking the aforementioned limitations of the model into account, we consider the geometry and orientation of the model molecular region to be reliable within the scope of this work. Owing to their nature, their uncertainties are not quantified in Table 1. The velocity field, on the other hand, is very reliable, with very small uncertainties. The reason for this is that this parameter is easy to measure in the high-spectral resolution ALMA maps, and quickly diverges from a fair fit when increased even slightly from the best-fit model. 

Temperatures are also well constrained, as a result of the fit to transitions with very different $J$, that is, with different excitation conditions, including some optically thick low-$J$ transitions, which probe temperatures better. This is especially true in the hottest structure, the main ring, whose emission is intense in single-dish profiles across all the $J$ examined in this work. Relative errors for the temperature in these regions are around 10\% with the simple method used in Table 1 for estimating uncertainties, although we consider a more conservative  error of around 20\% could be expected in those regions, given the density-temperature interplay. The relative error in the coldest structures, on the other hand, is larger, reaching 50\% in some structures. In any case, their best-fit temperature gives an idea of the characteristic temperatures found at those distances from the central star.

The total mass, which essentially depends on the total number of emitting molecules present, is well determined since we fitted transitions with different excitation conditions, four of which are optically thin. To this respect, the uncertainty in the total mass of the molecular region is not larger than the uncertainty of the \trece\ abundance, which we estimate at a factor 2.

Uncertainties in the density are more complex to deal with. The two simple laws used to reproduce the data  provide a fair global fit: constant density in the inner ring and inner
part of the main ring, and density falling with $r$ in the rest. Although the details of the specific distribution of the density are uncertain. Errors are lower in regions with strong emission, such as the inner part of the main ring, where they range between 8\% and 23\%. Errors in low-temperature regions affected by significant flux loss, instead, may reach up to 50\% or more. Actual errors could be somewhat larger because of the uncertainty in the thickness and geometry of the different structures. Also, the real structure could be very clumpy, in which case these model densities would be significantly underestimated.

Despite these caveats and the large uncertainties in the individual components of the model, the figures given for this model provide reliable characteristic physical conditions for the molecular  region of NGC~6302, even though their particular distributions might be not. 

\begin{table*}\renewcommand{\arraystretch}{1.3}
  \begin{center}
  \caption{Best-fit model physical parameters for the molecular region of NGC~6302. For the structure labelling and the geometric shape of each structure, refer to Fig.~\ref{Fsketch} or to the model file in {\tt SHAPE} format available as on-line supporting data.}
  \label{T11}
  \begin{tabular}{|l|c|c|c|c|c|c|}\hline 
{\bf Structure}  &  $\delta_\mathrm{V}$ & $n$ & $T$ &  $V_\mathrm{exp}$ & Age & M \\   
 &  {\scriptsize (km~s$^{-1}$)} & {\scriptsize (10$^{4}$~cm$^{-3})$} & {\scriptsize (K)}  &  {\scriptsize (km~s$^{-1}$)} & {\scriptsize (1000 yrs)} & {\scriptsize (\msun)} \\   
\hline
\hline
\multicolumn{4}{|l|}{{\bf Inner ring}} & \scriptsize{linear,$V_\mathrm{max}$} & & \\
\hline
IR    & 2 & 14$^{+7}_{-6}$ & 50$^{+15}_{-10}$ & 11$^{+0.5}_{-0.5}$ & 2.2 & 2.7$\times$10$^{-3}$  \\
\hline
\multicolumn{4}{|l|}{{\bf Main ring (inner region)}} & \scriptsize{constant} & & \\
\hline
\hline
MR-B1 & 1 & 35$^{+8}_{-6}$ & 260$^{+25}_{-20}$ & 8$^{+0.5}_{-0.5}$ & 3.5 - 3.8 & 1.1$\times$10$^{-2}$  \\
MR-R1 & 2 & 60$^{+5}_{-10}$ & 200$^{+20}_{-15}$ & 11$^{+1}_{-2}$ & 2.5 - 2.7 & 2$\times$10$^{-2}$ \\
\hline
\hline
\multicolumn{2}{|l|}{{\bf Main ring (outer region)}} & \scriptsize{$n$ range} & & \scriptsize{linear,$V_\mathrm{max}$} & & \\
\hline
MR-B2 & 2 & (16.2 -12.2)$^{+7}_{-8}$ & 40$^{+15}_{-15}$ & 13.1$^{+1}_{-1}$ & 2.9 & 4.7$\times$10$^{-2}$ \\
MR-R2 & 2 & (7.5 - 5.7)$^{+4}_{-3}$ & 40$^{+15}_{-15}$ & 10.4$^{+1}_{-1}$ & 3.7 & 1$\times$10$^{-2}$ \\
MR-R3 & 2 &  (0.9 - 0.7)$^{+0.1}_{-0.2}$ & 180$^{+30}_{-20}$ & 14.4$^{+1}_{-1}$ & 3.7 & 1.4$\times$10$^{-4}$ \\
\hline
\hline
\multicolumn{2}{|l|}{{\bf Lobe fragments}} & \scriptsize{$n$ range} & & \scriptsize{linear,$V_\mathrm{max}$} & & \\
\hline
LF-B1 & 2 & (2.9 - 1.5)$^{+1.4}_{-0.8}$ & 40$^{+20}_{-20}$ & 17$^{+2}_{-1}$ & 4.3 & 2.1$\times$10$^{-3}$ \\
LF-B2 & 2 & (5.6 - 3)$^{+2}_{-1.7}$ & 80$^{+30}_{-25}$ & 24.1$^{+2}_{-2}$ & 3.1 & 4.4$\times$10$^{-3}$ \\
LF-B3 & 2 & (2.1 - 1.4)$^{+0.8}_{-1}$ & 40$^{+15}_{-20}$ & 13.1$^{+1}_{-1}$ & 4.6 & 7.9$\times$10$^{-4}$ \\
LF-B4 & 2 & (2.1 - 1.3)$^{+0.9}_{-0.7}$ & 40$^{+15}_{-15}$ & 14.2$^{+1}_{-1}$ & 4.6 & 6.5$\times$10$^{-4}$ \\
LF-B5 & 2 & (1 - 0.35)$^{+0.4}_{-0.2}$ & 40$^{+20}_{-10}$ & 20.5$^{+2}_{-1}$ & 5.1 & 1$\times$10$^{-3}$ \\
LF-B6 & 2 & (4.3 - 3.3)$^{+2}_{-2.5}$ & 40$^{+15}_{-15}$ & 12.3$^{+1}_{-1}$ & 4.1 & 1.1$\times$10$^{-3}$ \\
LF-R1 & 2 & (0.7 - 0.5)$^{+0.3}_{-0.2}$ & 40$^{+15}_{-20}$ & 11.8$^{+1}_{-1}$ & 5.1 & 6.2$\times$10$^{-4}$ \\
LF-R2 & 2 & (1.7-1)$^{+0.7}_{-0.9}$ & 40$^{+20}_{-15}$ & 17.6$^{+2}_{-1}$ & 3.7 & 1.2$\times$10$^{-3}$ \\
\hline
\hline
  \end{tabular}
 \end{center}
\vspace{1mm}
 \scriptsize{
  {\it Parameters:}  \\
   $\delta_\mathrm{V}$ is  the characteristic microturbulence velocity of the structure. \\   
  $n$ is the density of the structure. When only one value is shown, the density is constant across the structure. When a range of values is shown, the density falls with the inverse of the distance to the star, and the values shown reflect the maximum and minimum density of the given structure. \\
  $T$ is the temperature of the structure.\\
  $V_\mathrm{exp}$ is the expansion velocity. In the case of the inner region of the main ring, it is a constant value. In every other case, the value of the velocity shown corresponds to the farthest edge of the structure from the star, with the velocity following a linear, ballistic expansion pattern. \\
  Age refers to the kinematical age of the structure, and $M$ to its mass.
}
\end{table*}

\subsection{Lifetime estimate of the optical lobes}

In order to better understand the evolution of the different ejecta of NGC~6302 and put them in context with each other, we tried to study the characteristic kinematic ages of the optical outflows in this nebula. Among the optical lobes, only the NW has been carefully analysed by \cite{meaburn08}, who estimated its kinematic age in 2200 years based on the apparent expansion of several knots of material in the period from 1956 to 2007. \cite{meaburn05} used {\tt SHAPE} to build a spatiokinematical model of the ionised component of NGC~6302 to match against a H$\alpha$+[N {\sc ii}] image and a set of echelle spectra taken with MES-SPM (\citealp{meaburn03}). These authors, however, only provided the parameters for the NW lobe, and did not provide the parameters of the wide-open lobes; this is probably because the only spectral information available sampling those lobes was in a single echelle spectrum passing through the main axis of the nebula (position 1), and therefore the uncertainties would have been too large to draw any solid conclusions.

Keeping that in mind, we derived a rough estimate of the kinematic age of these wide-open lobes by means of a  {\tt SHAPE} spatio-kinematical model. We simultaneously fitted the F658N HST image and the echelle spectrum analysed by \cite{meaburn05}. This allowed us to loosely constrain the inclination of the structure between 12 and 17$^\circ$, and to constrain its kinematic age between 3,600 and 4,700 years at the adopted distance of 1.17~kpc. Nevertheless, owing to the uncertainty of the model, we stress that these figures should be taken as a rough estimate.

\section{Discussion}

It is becoming increasingly clear that the main agents in shaping PNe operate at its innermost region, where a significant equatorial density enhancement should be present and related to the collimation of light and jet-launching from the central star preferentially towards the polar directions (e.g. \citealp{balick02}). Most of the material forming this equatorial condensation must be lost by the AGB star as wind deployed in the circumstellar medium. Alternatively, the presence of a potential companion star would gather part of this material on a rotating disk, releasing an angular momentum excess through magnetic forces, and thus powering the jets, in a similar process as that at work in young stars (e.g. \citealp{soker01}, \citealp{frank04}). Such circumstellar disks, with proposed sizes of a few stellar radii, have never been directly observed, although larger, circumbinary orbiting disks have indeed been found in a number of post-AGB stars (e.g.~\citealp{bujarrabal13b}); some of these disks were even resolved (the Red Rectangle, \citealp{bujarrabal05,bujarrabal13c,bujarrabal16}; AC~Her, \citealp{bujarrabal15}; and IW~Car, Bujarrabal et al. in preparation). Whether this equatorial density enhancement precedes the shaping of the bipolar nebulae, as theoretical models predict for orbiting disks, or is coeval (thus not being the direct cause of nebula shaping), is something yet to be proven. In any case, these compact equatorial regions are predicted by the models and are directly related to the collimation of light and outflows from the star.

The innermost region of NGC~6302 shows a central gap created by the photodissociation region (PDR), which is inferred from the presence of atomic carbon (both neutral and ionised) with a kinematic distribution that is compatible with its location near the CO-rich region, but closer to the star (\citealp{santander15a}). The ALMA data analysed in this work, however, show no signature of a circumstellar or circumbinary flat disk in Keplerian rotation, thus effectively putting an upper limit on its size: should such a CO-rich disk exist and be effectively shielded from the UV radiation from the central star, then assuming a brightness temperature of 50~K  (as in the inner ring and the rotating disk found the Red Rectangle, \citealp{bujarrabal13}), its diameter must be $\lesssim$100 UA, or 1.5$\times$10$^{15}$~cm, for its emission to remain undetected with the sensitivity attained by these ALMA observations.

Concerning larger scale structures, the physical conditions of the main ring found in this work are compatible with previous findings, with slightly larger densities and temperatures, a lower \doce and \trece abundance than found by \cite{santander15a}, and a similar total molecular mass of 0.1 \msun. Our modelling has also allowed us to derive kinematical ages of the different structures, and thus consider the following sequence of events in the ejection of the nebula of NGC~6302: the bulk of the material was ejected and shaped into a ring of material and a set of lobe fragments in an event starting 5,000 years ago and lasting for about 2,000 years. The optical lobes were ejected after a brief delay, at some uncertain point between 3,600 and 4,700 years ago, followed by the large and fast-expanding northwest lobe that was ejected around 2,200 years ago (\citealp{meaburn08,szyszka09}). The new, inner ring was ejected approximately at the same time as the NW lobe, 2,200 years ago. The star left the AGB stage some 2,100 years ago according to photoionisation models and evolutionary tracks by \cite{wright11}, so the inner ring was presumably ejected in the last stages of the transition to PNe. 

In other words, NGC~6302 seems to have undergone two distinct episodes of nebular ejection, with the formation of the main ring and the ejection of the perpendicular optical lobes first, and the NW lobe and inner ring ejected several hundreds of years later (see Sect. 4.2). No association between the latter two is however clear in the data. The NW lobe runs roughly perpendicular to the main ring, rather than to this new structure, although the projection of its axis in the plane of the sky seems to run approximately along the major axis of the ellipse which traces the inner ring in the plane of the sky. In any case, the conclusion to be drawn from the presence of these seemingly associated structures remains a mystery, although it allows for some further discussion.

During the ejection sequence described above, the molecular gas has been presumably swept by newer ejecta or winds from the central star, and photodissociated to form a very clumpy molecule-rich medium. The velocity of the receding part of the main ring is largest in its inner rim, resulting in a much shorter kinematic age of 2,500-2,700 years. This constitutes evidence of the presence of shocks pushing away and accelerating the previously deployed material, as suggested by \cite{santander15a}, and indicates that the actual age of the inner rim of that part of the main ring is somewhat larger than what the kinematic age predicts. These shocks in the receding part of the nebula might also have affected the northwestern region of the inner ring (most of which is also receding), breaking it into a set of irregular clumps. 

The molecular region tends to expand faster in the southern region, around where the main ring is broken or  missing (see Fig.~\ref{Fsketch}). In particular, the kinematic age of the structures in that direction, labelled as MR-R3 and LF-R2, are 3,700 years; this is a considerably younger age than most of the lobe fragments at those distances from the star. In principle, the lack of emission from this southern region of the ring could be related either to a sudden drop of density or to faster photodissociation by the star. We performed an abundance contrast test and found that the CO density in the affected region must be 150 times lower than in the rest of the ring in order for the emission to be at the noise level in the optically thin \trece map. Such a large density drop in such a small region is unlikely, so we conclude that the ring is broken because of the destruction of CO by photodissociation or shocks. This supports the findings by \cite{uscanga14}, whose 3D hydrodynamic model suggests that an additional acceleration mechanism, such as the rocket effect due to photoevaporation of CO clumps, is acting near the central star. To this respect, \cite{szyszka09} find\ evidence of acceleration  of the gas in the inner regions by an additional 9.2 \kms, which relates to the onset of ionisation.

\subsection{The inner ring}

The inner ring deserves further discussion. It is inclined with respect to, and is considerably younger than, the main ring. Its spatial location coincides with the arc filament identified by \cite{matsuura05} in a HST H$\alpha$ image in their Fig.~8, with the ionised emission from the optical image seemingly tracing the inner rim of this structure (see Fig.~\ref{Fmoments}). This proves both features to be manifestations of the same structure expanding at $\sim$11 \kms outwards from the star. The orientation of this inner ring differs from that of the main ring by $\sim$60$^\circ$; the axis of the inner ring does not lie far from the direction of the missing section of the latter (see the bottom right panel in Fig.~\ref{Fsketch}) and, based only in HST imagery, is approximately perpendicular to the projection onto the sky of the cirrus structures visible in the innermost region of the north-east outflow depicted by \cite{matsuura05} in their Fig.~7. However, the inner ring is not perpendicular to the whole extent of that structure, which progressively curves towards the east. The mass we find for this structure is 2.7$\times$10$^{-3}$ \msun.

Multiple rings in expansion close to the nuclei of PPNe or PNe have been reported in the literature. At an adopted distance of 3~kpc, Frosty Leo shows a central ring-like structure with a similar radius of 7.2$\times$10$^{16}$~cm, inclined $\sim$42$^{\circ}$ with respect to the main axis of the optical nebula. It expands three times faster than the ring discussed in this work and shows fast-expanding polar jets (\citealp{castro05}). The Frosty Leo ring, however, is far more massive than the ring in NGC~6302, amounting up to 0.8 \msun,  thus constituting the bulk of the molecular envelope of Frosty Leo. The origin and dynamics of the Frosty Leo ring were attributed by the authors to multiple post-AGB interactions, in which ejections after  the optical lobes bored holes in other directions, modifying the geometry of the inner nebular regions. M~2-9 also has a 2,100 year old molecular ring expanding at $\sim$7 \kms, of a similar size with a radius of 4.5$\times$10$^{16}$~cm, as found by \cite{zweigle97},  adapted for the distance of 1,320 pc and recomputed by \cite{corradi11b} after finding an error in the (still) widely-used determination by \cite{schwarz97}. A second ring with a tighter radius of 1.8$\times$10$^{16}$~cm (for the same distance) was found by \cite{castro12} expanding at half the velocity of its larger counterpart. These rings, however, preceded the ejection of the optical lobes by hundreds of years and are essentially aligned with them, whereas the mass of the larger ring is at least 9$\times$10$^{-3}$~\msun, of the same order as the ionised mass of the whole nebula (\citealp{kwok85}).

While the mass of the main molecular ring of NGC~6302 is of the same order as the 0.1-0.4 \msun\ ionised mass of the nebula (\citealp{huggins89,huggins96}, adapted to the distance used in this work), the mass of the innermost ring is two orders of magnitude lower (comparable to the mass of a gaseuous, 2.8 M$_J$ giant planet), and, remarkably, the structure is as young as the NW optical lobes themselves. \cite{soker16} provides an exotic hypothesis, which could account for the presence of both rings and involves a tight binary system in orbit around an AGB star. In this triple-stellar evolution scenario, the tight binary system could break up in the outskirts of the envelope of the AGB, producing two distinct episodes of equatorial mass ejections that can lie in different planes, as is indeed the case in NGC~6302.

It is tempting, although highly speculative with the data at hand, to also consider the possibility that the inner ring is the result of the destruction, evaporation, and ejection of a gas giant planet of a few Jupiter masses. Planetary ingestion is expected to happen on the AGB if a planet is initially orbiting within $\sim$10 AU of the star in the main sequence (\citealp{nordhaus10,nordhaus13}). \cite{villaver08} and \cite{villaver11b} find that a planet could survive the AGB and yet be close enough to the star so as to be evaporated. The evaporation rate would depend on the planet's surface temperature, which would greatly increase as the luminosity and effective temperature of the central star rises as it evolves towards the PN stage, after deploying the bulk of its molecular envelope. In the case of NGC~6302, the evaporation rate should have been as large as 10$^{-4}$ o 10$^{-3}$ M$_\mathrm{J}$ yr$^{-1}$ in order for the material to be evaporated along a few orbits, and ejected into a relatively tight ring. In any case, further theoretical simulations would be necessary to address such a speculative possibility.

\section{Conclusions}

The main results of this work are summarised in what follows. We present ALMA band 7 observations of the dense molecular region of NGC~6302. The general structure consists of a broken ring out of which a system of faint, curved lobe fragments emerge, running parallel or enclosing the optical lobes. There is no indication of a Keplerian disk at the innermost region around the central star, which, if it exists, must have a diameter smaller than 100 UA at the adopted distance of 1.17~kpc. We detect the \doce\ and \trece \jtd\ transitions, as well as fainter emission lines from SO and NS.

Our modelling recovers the geometry of the ejecta in great detail, providing a good fit to both the ALMA maps and previous data from CO \jtd\ single dish data from JCMT and high-$J$ transitions observed with HERSCHEL/HIFI. The velocities, densities, and temperatures we find are similar to previous models by \cite{dinhvtrung08} and \cite{santander15a}, including abrupt velocity changes indicative of shocks, although the derived \doce\ and \trece\ abundances are  1$\times$10$^{-4}$ and 1.25$\times$10$^{-5}$, respectively, which is a factor 1.5 and 2 lower than in models by those authors.

We detect and model a previously unreported structure consisting of a thin, inner ring with an ionised arc-like feature being its counterpart in optical images. The ring is considerably inclined with respect to the larger molecular ring, has an average radius of 7.5$\times$10$^{16}$~cm, a mass of 2.7$\times$10$^{-3}$ \msun\ , and expands at 11 \kms. The latter implies a kinematical age for this ring of just $\sim$2200 years, therefore being ejected after the bulk of the molecular region, and approximately at the same time as the NW lobes visible in the optical.

\begin{acknowledgements}
MSG wishes to thank Eva Villaver for her insight on the part of the discussion dealing with planet engulfment and ejection, and John Meaburn, the referee, who contributed to improve this work. This paper makes use of the following ALMA data: ADS/JAO.ALMA\#2013.1.00331.S. ALMA is a partnership of ESO (representing its member states), NSF (USA) and NINS (Japan), together with NRC (Canada), NSC and ASIAA (Taiwan), and KASI (Republic of Korea), in cooperation with the Republic of Chile. The Joint ALMA Observatory is operated by ESO, AUI/NRAO, and NAOJ. This paper makes use of data from Herschel/HIFI. Herschel is an ESA space observatory with science instruments provided by European-led Principal Investigator consortia and with important participation from NASA. This work was partially supported by the Spanish MINECO within the programme ``F\'\i sico-qu\'\i mica del medio interestelar y circunestelar en la era de ALMA'' (AYA2012-32032), and by the European Research Council (ERC Grant 610256: NANOCOSMOS).
\end{acknowledgements}



\bibliographystyle{aa}
\bibliography{msantander}


\newpage
\appendix

\section{Emission from SO and NS}

\begin{figure}[!h]
\center
\resizebox{10cm}{!}{\includegraphics{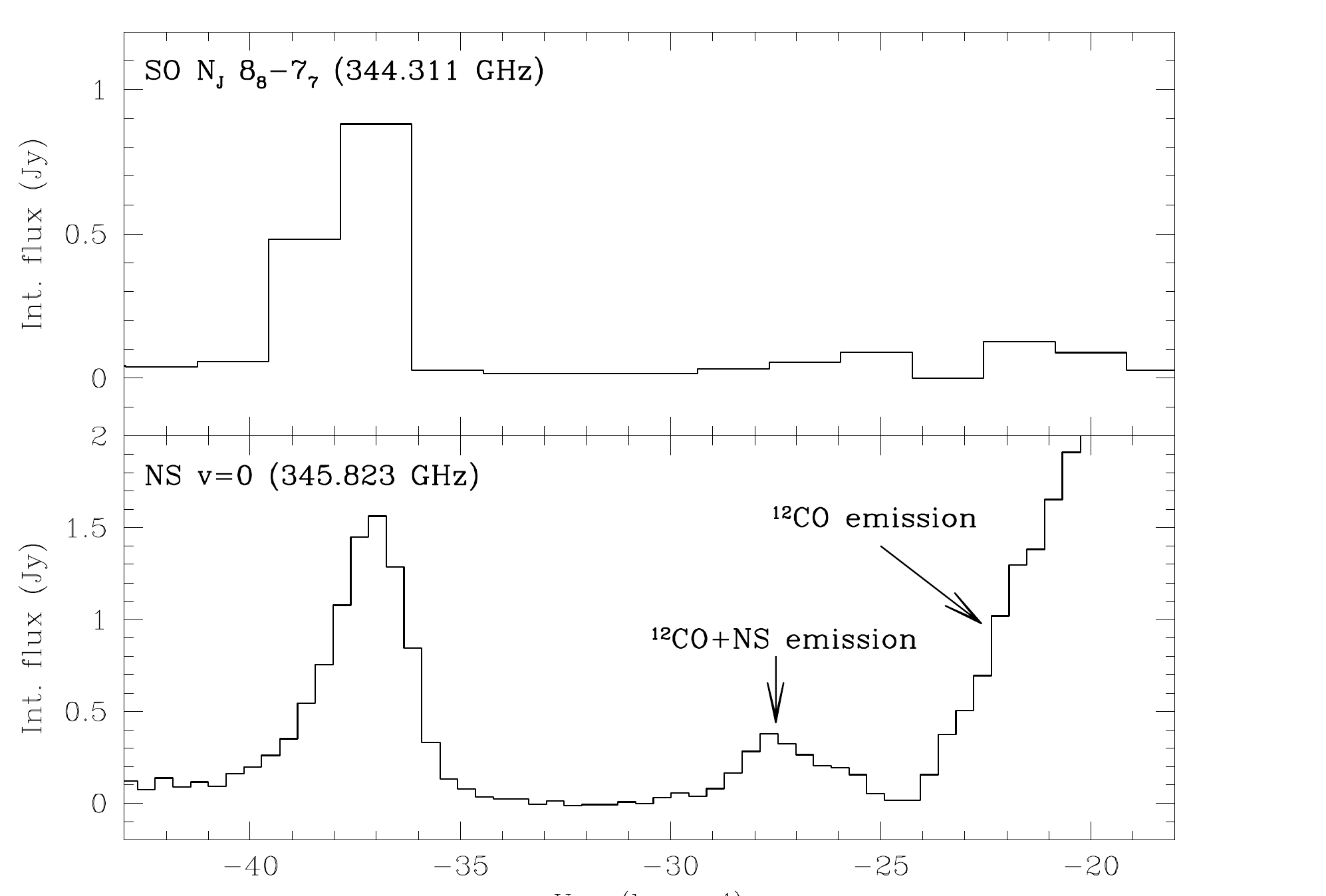}}
\caption{{\bf Top panel:} Continuum-subtracted, spatially-integrated emission from the N$_\mathrm{J}$, 8$_8$-7$_7$ line of SO (344.311 GHz). {\bf Bottom panel:} Continuum-subtracted, spatially-integrated emission from NS (345.823 GHz). The emission from NS partially overlaps with that from \doce\ \jtd\ (in Fig.~\ref{Flost12}, the emission around \vlsr=-62 \kms\ corresponds to NS).}
\label{Fothermol}
\end{figure}

\begin{figure*}[!]
\center
\resizebox{15.6cm}{!}{\includegraphics{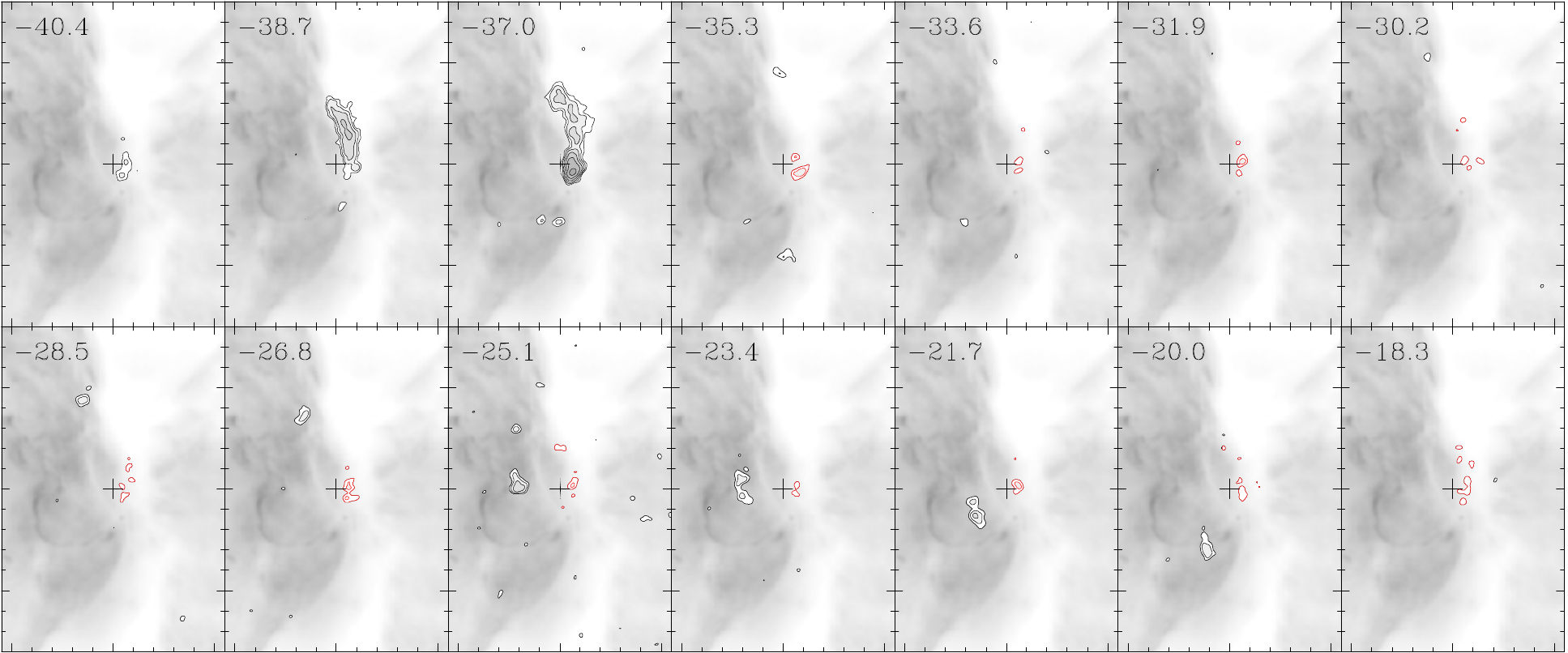}}
\caption{ALMA N$_\mathrm{J}$, 8$_8$-7$_7$ SO emission map of NGC~6302. The field of view is 11$''\times16''$. \vlsr\ velocities in \kms\ are indicated in each channel. The intensity scale of the contours is logarithmic, with the first contour at 10 \mjyb and a step of 2 dB (i.e. a factor of 10$^{0.2}=1.585$) between adjacent contours. The continuum has not been subtracted, and can be seen owing to the low intensity of the contours highlighted in red. An HST F673N image of the central region of NGC~6302 is shown in the background for comparison purposes.}
\label{FmapSO}
\end{figure*}

The permitted N$_\mathrm{J}$, 8$_8$-7$_7$ line of SO (344.311 GHz) is visible in the spectral window corresponding to the continuum near \doce \jtd\ (HPBW 0\farcss 4$\times$0\farcss 34, major axis along PA 102$^\circ$). Its spatially-integrated emission is shown in Fig.~\ref{Fothermol}. The ALMA map is shown in Fig.~\ref{FmapSO}. The extended spatial distribution is coincident with the region where the main ring reaches its brightness peak in \doce\ and \trece\ \jtd, that is, at $\sim$-37 \kms, although the top rim of the ring is also (barely) visible at $\sim$-27 \kms, as is the inner rim of the opposite side of the ring at $\sim$-21 \kms. The emission traces the main ring, although the contrast between the approaching and receding parts is much larger than in \doce\ and \trece\ \jtd\ (see their integrated emission in Figs.~\ref{Flost12} and \ref{Flost13}), thus rendering the latter very faint. The integrated line profile of SO peaks at  0.88 Jy above the continuum, while its velocity-integrated emission is 3.3 Jy~\kms. 

\begin{figure*}[!]
\center
\resizebox{15.6cm}{!}{\includegraphics{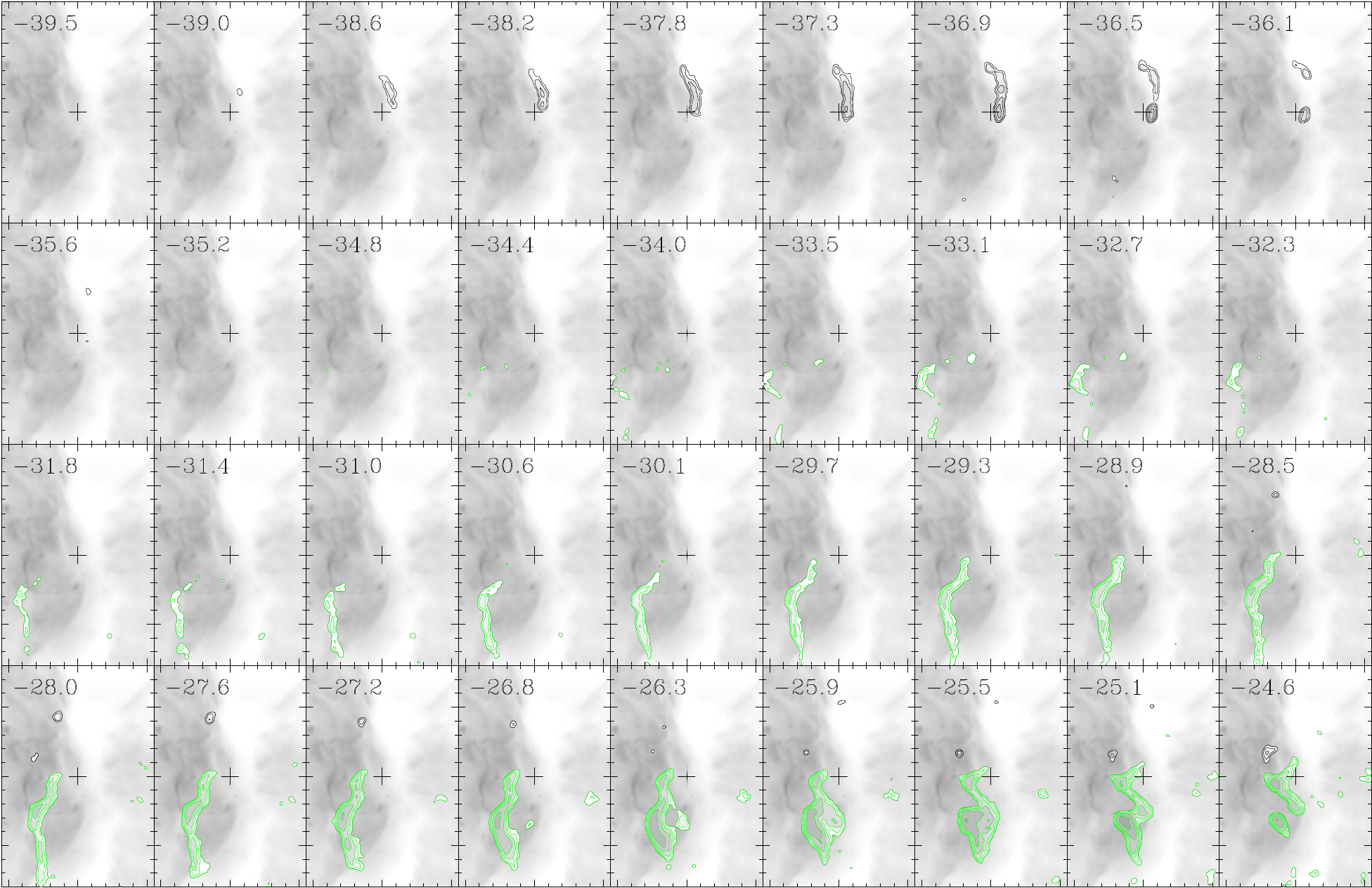}}
\caption{ALMA NS emission map of NGC~6302. The field of view is 11$''\times16''$. \vlsr\ velocities in \kms\ are indicated in each channel. The intensity scale of the contours is logarithmic, with the first contour at 25 \mjyb, and a step of 2 dB (i.e. a factor of 10$^{0.2}=1.585$) between adjacent contours. Severe contamination from \doce\ \jtd\ is apparent  from \vlsr\ -34.4 \kms\ to -24.6 \kms, and was highlighted here in green contours. An HST F673N image of the central region of NGC~6302 is shown in the background for comparison purposes.}
\label{FmapNS}
\end{figure*}

Emission from NS (345.823 GHz) is also clearly detected in the spectral window corresponding to \doce \jtd, also peaking at $\sim$-37 \kms. Its spatially integrated emission is shown in Fig.~\ref{Fothermol}. The corresponding ALMA map is shown in Fig.~\ref{FmapNS}. This emission traces the main ring, which also shows a large contrast between the approaching and receding parts in a similar fashion to the emission from SO. Emission from this receding part in NS partially overlaps with emission from \doce\ at large velocities in the range \vlsr\ -50 to -45 \kms. Its integrated line profile peaks at 1.5 Jy, while its velocity-integrated emission (computed in the non-overlapping region of the spectral profile) amounts to 5.1 Jy~\kms. 

Both SO and NS are molecules expected in O-rich envelopes such as that of NGC~6302 (e.g. OH231.8+4.2; see \citealp{sanchez14}).

\section{Emission from \trece\ \jtd\ and model}

\begin{figure*}[!]
\center
\resizebox{15.6cm}{!}{\includegraphics{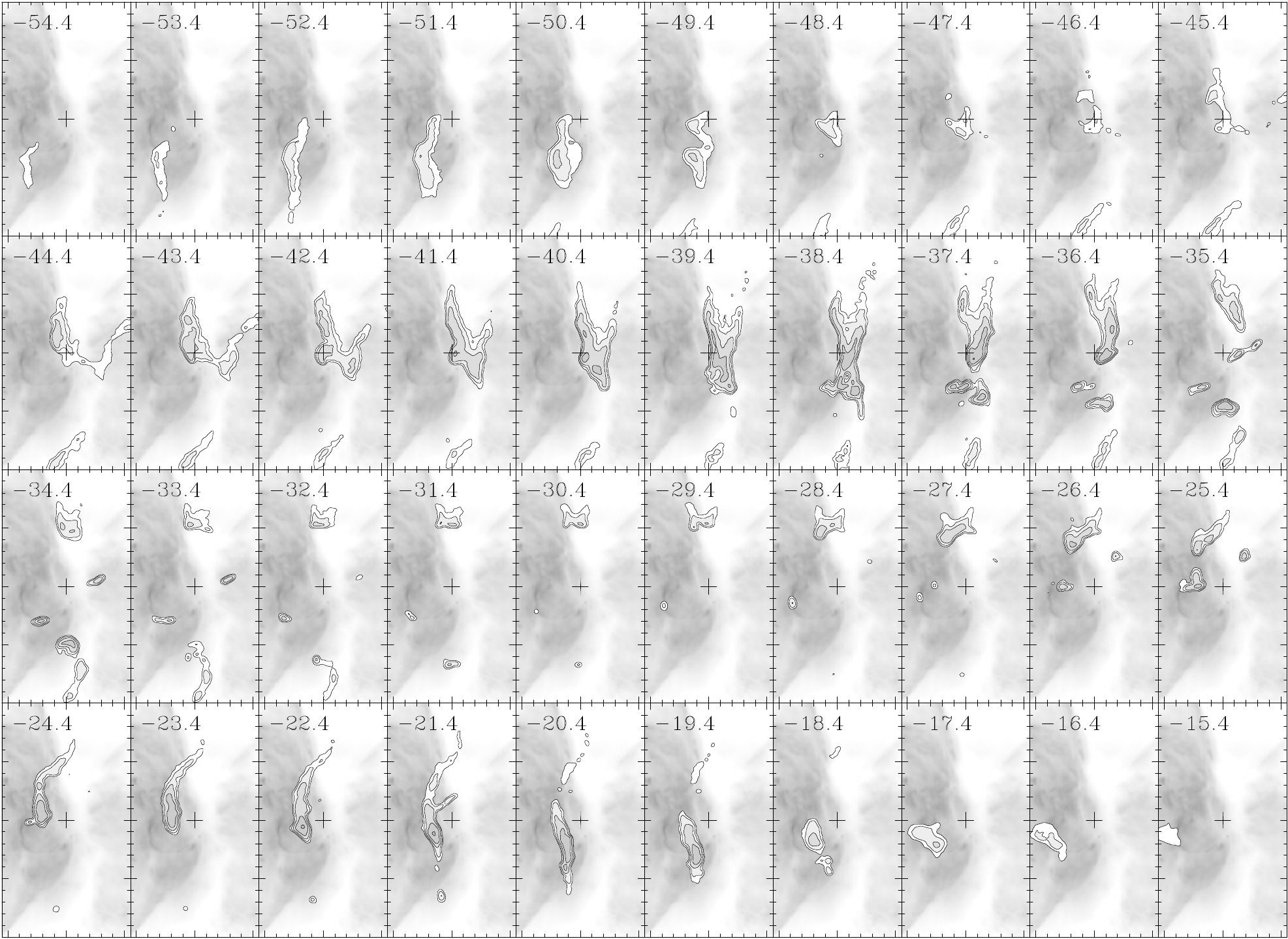}}
\caption{Same as Fig.~\ref{F12obs}, but for \trece\ \jtd\ emission. The first contour is at 20 \mjyb, with a step of 3 dB between adjacent contours. The deepest negative contour has a level of -30 \mjyb.}
\label{F13obs}
\end{figure*}

\begin{figure*}[!]
\center
\resizebox{15.6cm}{!}{\includegraphics{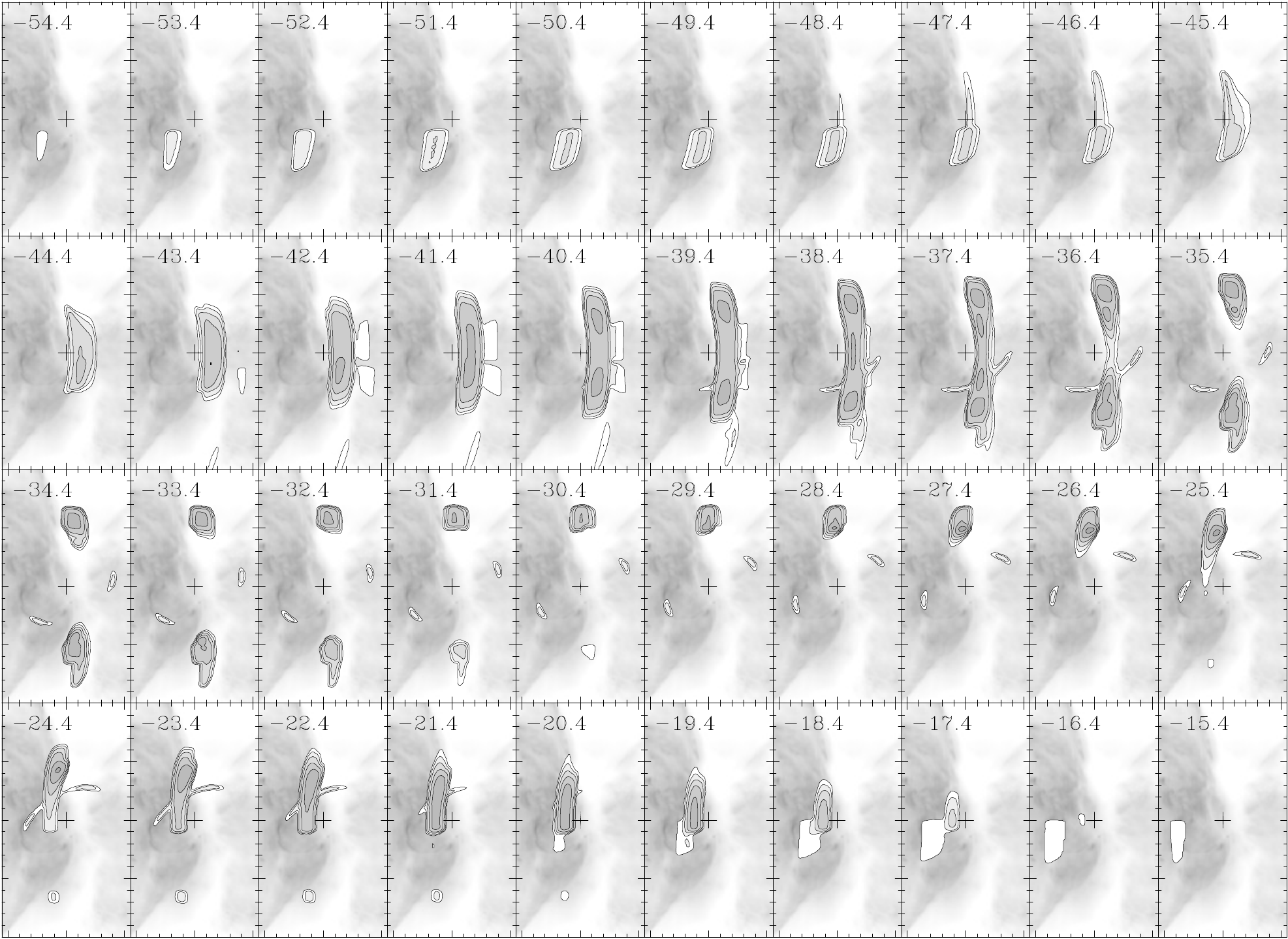}}
\caption{Same as Fig.~\ref{F12mod}, but for \trece\ \jtd\ emission. The first contour is at 20 \mjyb, with a step of 3 dB between adjacent contours.}
\label{F13mod}
\end{figure*}

The observations and model of the \trece\ \jtd\ emission of NGC~6302 are shown in Figs.~\ref{F13obs} and \ref{F13mod}.

\end{document}